\documentclass[prd,aps,tightenlines,superscriptaddress,nofootinbib,preprint]{revtex4}

\usepackage{graphicx}
\usepackage{dcolumn}
\usepackage{bm}
\usepackage{color}
\usepackage{amsmath}
\usepackage{amsfonts}
\usepackage{amssymb}
\usepackage{orcidlink}
\begin{document}


\title{Physical constraints on the Maldacena-Shenker-Stanford chaos-bound in black hole spacetimes}

\author{Terkaa Victor Targema \orcidlink{0000-0001-8809-1741} \footnote{{\color{purple}s2571502@ipc.fukushima-u.ac.jp,~terkaa.targema@tsuniversity.edu.ng}}}
\affiliation{Faculty of Symbiotic Systems Science, Fukushima University, Fukushima 960-1296, Japan}
\affiliation{Department of Physics, Taraba State University, Jalingo, Nigeria}
\author{Kazuharu Bamba \orcidlink{0000-0001-9720-8817}
\footnote{{\color{purple}bamba@sss.fukushima-u.ac.jp}}}
\affiliation{Faculty of Symbiotic Systems Science, Fukushima University, Fukushima 960-1296, Japan}

\author{Riasat Ali \orcidlink{0000-0002-9371-1113} \footnote{{\color{purple}riasatyasin@gmail.com}}}
\affiliation{Department of Mathematics, GC
University Faisalabad Layyah Campus, Layyah-31200, Pakistan}

\author{Usman Zafar \orcidlink{0000-0001-9610-1081} \footnote{{\color{purple}s2471001@ipc.fukushima-u.ac.jp,~zafarusman494@gmail.com}}}
\affiliation{Faculty of Symbiotic Systems Science, Fukushima University, Fukushima 960-1296, Japan}


\begin{abstract}
 Chaotic motion near black holes has recently been examined through the lens of the Maldacena-Shenker-Stanford (MSS) chaos-bound, but reported violations remain contradictory. A significant source of ambiguity stems from treating the particle angular momentum as an independently adjustable parameter instead of as a quantity fixed by the circular-orbit conditions. We develop a constrained framework in which the angular momentum is determined self-consistently from the geometry. Applied to the charged Kiselev black hole, this framework shows that certain previously reported violations of the chaos bound can be attributed to inconsistent parameter choices rather than to intrinsic curvature effects. By extending the analysis to geometries containing higher-order curvature terms, we find genuine chaos-bound violations at large charge-to-mass ratios, originating from curvature corrections rather than orbital parameters. Our approach, therefore, provides a systematic means to distinguish between parameter-induced (apparent) and curvature-induced (physical) violations in Einstein gravity and its extensions.

\end{abstract}
                             
\maketitle


\section{Introduction}
The Lyapunov exponent \(\lambda\) quantifies the exponential sensitivity of a dynamical system to initial conditions and serves as a standard diagnostic of chaos in classical mechanics. In quantum many-body systems, chaotic behavior is characterized using out-of-time-ordered correlators (OTOCs), which probe the growth of operator commutators at different times \cite{A1, A2, A3, A4, A5}. In chaotic regimes, these correlators exhibit exponential growth,
\(
C(t) \sim e^{\lambda t},
\)
with \(\lambda\) identified as the quantum Lyapunov exponent. A breakthrough result by Maldacena, Shenker, and Stanford \cite{maldacena2016bound} established a universal upper bound on the Lyapunov exponent for large-$N$ thermal quantum systems, given by
\begin{equation}
    \lambda_L \le \frac{2\pi k_B T}{\hbar}\,,
\end{equation}
under assumptions of analyticity, causality, and an exponential OTOC growth regime, where $T$ is the temperature, $k_{B}$ is the Boltzmann constant, and $\hbar$ is the reduced Planck constant. Within gauge/gravity duality (AdS/CFT), thermal states of certain strongly coupled quantum systems admit a classical description in terms of AdS black holes. At large $N$ and strong coupling, with probes treated in the semiclassical limit, boundary OTOC growth can be traced to bulk near-horizon dynamics: controlled shockwave analyses demonstrate that the boundary Lyapunov exponent is determined by the surface gravity $\kappa$ of the dual horizon \cite{shenker2014black, wang2022pole}.

With consistent bulk–boundary time matching, the OTOC Lyapunov exponent acquires a classical bulk interpretation and coincides with the instability exponent of suitable probes \cite{pourkhodabakhshi2025saturation}. For holographic black hole backgrounds, the relevant instability originates from the universal near-horizon exponential redshift and is therefore fixed by $\kappa$ (in units with $\hbar = k_B = 1$). Unstable circular orbits sufficiently close to the horizon, accordingly, exhibit a Lyapunov exponent that approaches $\kappa$; see Refs.~\cite{hashimoto2017universality, gwak2022violation, lei2021chaos, prihadi2023chaos} for detailed analyses. These results motivate the geometric bound
\begin{equation}
\lambda_{\rm probe} \le \kappa\,,
\end{equation}
to be interpreted as the bulk (geodesic/probe) counterpart of the MSS bound.

A particularly controlled framework arises when particles are subjected to external forces, allowing them to probe arbitrarily close to the horizon without crossing it \cite{copson1928electrostatics, hanni1973lines}. In Ref.~\cite{hashimoto2017universality}, the resulting near-horizon dynamics yield a Lyapunov exponent equal to the surface gravity, saturating the bound. Angular momentum generates homoclinic orbits away from the horizon but does not induce violations. Exceeding the bound requires nonstandard external interactions, such as higher-spin forces, whose modified near-horizon behavior alters the effective potential and enhances the instability rate.

However, recent studies reporting violations of the chaos bound—often attributing a central role to angular momentum—have continued to appear in the literature. For example, violations were reported in Einstein-Euler-Heisenberg AdS black holes for sufficiently large values of the angular momentum, charge, and Euler--Heisenberg parameter \cite{d7}.
 Similarly, in the case of both non-extremal and extremal Reissner–Nordstr\"om black holes with scalar hair, Ref.~\cite{wang2023spatial} identified spatial regions where violations could occur due to the combined influence of angular momentum, charge, and the location of circular orbits. Additionally, in the charged Kiselev black hole, chaos-bound violations arise predominantly at large angular momentum, with the normalization factor of the matter state parameter being a major contributing factor as discussed in Ref.~\cite{gao2022chaos}. More recently, it has been found that violations for \(p\)-branes, which in that case appeared to be independent of angular momentum and charge \cite{c2}. Several other examples of such violations can be found in Refs.~\citep{addazi2021chaotic, xie2023circular, c5, d2, d3,  lee2025bound, jeong2023homoclinic, d6, d8, d9, kan2022bound, gwak2022violation, lei2024thermodynamic, fayyaz2025chaotic}.  

Despite the extensive studies on this topic, the mechanisms responsible for the reported chaos-bound violations remain incompletely clarified. Understanding the circumstances under which such violations appear is important for assessing the robustness of the gravitational analogue of the MSS bound. In particular, it is necessary to formulate a framework applicable to generic spherically symmetric spacetimes in which the dynamical conditions defining circular motion are implemented consistently. This allows one to distinguish apparent violations arising from inconsistent orbital prescriptions from those that persist under fully imposed circularity conditions. In this work, we revisit this issue by examining how the interplay between angular momentum and charge governs the stability of circular orbits and the emergence of large Lyapunov growth rates. In particular, analyses based on the effective potential approach typically impose well-defined constraints on the angular momentum to ensure orbital consistency \cite{gwak2022violation, ali2025evaluation}. In contrast, within the Jacobi framework, it is often difficult to determine both the angular momentum and the corresponding circular orbits simultaneously. As a result, angular momentum is sometimes treated in numerical studies as an external parameter (\(L = 0, 5, 15, \cdots\)), which can be misleading particularly for circular orbits. Here we take the alternative perspective that $L$ for circular orbits should be determined self-consistently from the geometry. Using this constraint we demonstrate that certain apparent violations previously reported arise from this methodological difference, while real violations occur and can be linked to curvature corrections beyond the geometric constraints.
Therefore, by extending the Jacobi formalism, we try to develop a consistent analytical framework for any spherically symmetric spacetime where angular momentum ($L$) and circular orbit ($r_{0}$) are determined together from the same dynamical structure. This framework maintains the consistency in all equilibrium conditions and offers a one-to-one mapping $r_{0}\leftrightarrow L$, linking the orbital radius and angular momentum.  This allows the Lyapunov exponent and the chaos-bound to be determined without any arbitrary assumptions about parameters, maintaining physical consistency. This approach simplifies the analysis, thereby providing a more transparent and physically consistent interpretation of the conditions under which real or apparent violations of the chaos-bound may arise.

To concretize our analysis, we examine two representative spacetimes: the charged Kiselev black hole surrounded by a cloud of strings and a cosmological constant, and the \(f(R)\)-modified black hole.  The distinction between these two BH models reveals a fundamental difference between constraints arising from geometry and those originating from curvature modifications. For example, in Kiselev spacetime, the geometric stability conditions that determine orbital angular momentum maintain stability against the influences of quintessence, string cloud, and charge effects, whereas, in \(f(R)\) gravity, higher-order-curvature terms yield corrections to the gravitational potential that lie beyond these geometric constraints. Therefore, these examples may offer a sufficiently broad parameter space to test the robustness of our formalism and to facilitate meaningful comparison with previously reported results in the literature.

Additionally, we demonstrate that the observed chaos-bound violations arise from genuine curvature effects rather than apparent effects of probe dynamics, thereby enhancing the conceptual clarity of quantum chaos bounds in gravitational systems. It is essential to clarify that the motion along circular geodesics in a static spherically symmetric spacetime remains integrable, meaning it can not be considered as deterministic chaos within classical dynamical-system theory. We do not treat the Lyapunov exponent as demonstrating classical non-integrable dynamics in this work; rather, it describes the exponential sensitivity of circular motion to minor deviations from equilibrium. In holographic frameworks, these instability rates are linked to the exponential growth of OTOCs, which are widely used to probe quantum chaos in the dual theory. Therefore, our comparison with the MSS bound should be considered an examination of instability growth in gravitational settings (purely geometrical interpretation of MSS bound instability), rather than an examination of classical chaos itself. This differentiation is crucial for properly understanding the physical meaning of our findings. The paper is organized as follows. Section~\ref{sec2} reviews the theory of Lyapunov exponents in black hole spacetimes.
We present the Jacobi-matrix formulation for static, spherically symmetric geometries, with supplementary details.
Additionally, we also introduce the effective potential approach as a complementary method for evaluating orbital stability. The main analysis in Sec.~\ref{san} applies these formalisms to the Reissner--Nordstr\"om and Kiselev black holes, as well as to charged black holes in \( f(R) \) gravity. Finally, Sec.~\ref{con} summarizes our main results and provides a possible future extension of our work.
\section{Theory of Lyapunov exponents}
\label{sec2}
The Lyapunov exponent quantifies how rapidly nearby trajectories in phase space diverge, making it a significant indicator of dynamic instability. From the perspective of curved spacetime, this technique offers a way to examine geodesic or orbital stability, providing a window into the connection between classical orbital dynamics and quantum chaos diagnostics. Therefore, in this section, we discuss the generalized framework for deriving Lyapunov exponents in curved spacetimes, which has been extensively developed in the literature. For the Jacobi method, which we briefly review in this section, further details can be found in Refs.~\cite{gallo2025bounds, c2, gao2022chaos, gwak2022violation, singh2025circular}. Let \(X_{i}(t)\) denote the state of a dynamical system at time \(t\) satisfying
\begin{equation}
\nonumber
\frac{dX_i}{dt} \;=\; F_i\bigl(X^j\bigr)\,,
\end{equation}
with initial condition \(X(0)=X_0\).  For a reference trajectory (e.g., a circular geodesic), small deviations \(\delta X_i(t)\) obey the linearized system
\begin{equation}
\nonumber
\frac{d}{dt}\,\delta X_i \;=\; K_{ij}(t)\,\delta X_j\,,
\qquad
K_{ij}(t)\;=\;\left.\frac{\partial F_i}{\partial X_j}\right|_{X(t)}\,.
\end{equation}
Writing
\begin{equation}
\nonumber
\delta X_i(t) \;=\; L_{ij}(t)\,\delta X_j(0)\,,
\end{equation}
the evolution matrix \(L_{ij}(t)\) satisfies
\begin{equation}
\nonumber
\dot L_{ij} \;=\; K_{ik}(t)\,L_{kj}\,,
\quad
L_{ij}(0)=\delta_{ij}\,.
\end{equation}
The Lyapunov exponent \(\lambda\) is then extracted from the late-time growth of the dominant eigenvalue of \(L_{ij}\). The representation of $\lambda$ is given by
\begin{equation}
\label{eqn: lambda_matrix}
\lambda = \lim_{t \to \infty} \frac{1}{t} \ln \|L(t)\|\, .
\end{equation}
In a quantum setting, chaos is often diagnosed by the OTOCs of two Hermitian operators \(V(0)\) and \(U(0)\), defined by
\begin{equation}
C(t)\;=\;-\,\bigl\langle [\,V(t),\,U(0)\,]^2 \bigr\rangle\,.
\end{equation}
In chaotic regimes, the relationship between OTOCs and the Lyapunov exponent can be expressed as  
$C(t)\;\sim\;e^{2\lambda\,t}.$
 In the semiclassical limit, commutators reduce to Poisson brackets,
\(
[\,V(t),\,U(0)\,]\;\longrightarrow\;i\hbar\,\{\,V(t),\,U(0)\,\},
\)
so that the exponential growth of OTOCs matches the classical instability rate stated in Eq.~(\ref{eqn: lambda_matrix}) (for more details, check Ref.~\cite{gwak2022violation}).
\subsection{Lyapunov exponent in spherical symmetric static spacetimes: Jacobi approach} \label{sec2a}
Consider the circular motion of a charged particle around a spherically symmetric black hole. The line element is typically of the form
\begin{equation}
\label{ds}
ds^2 = -f(r) dt^2 + \frac{1}{h(r)} dr^2 + B(r) d\theta^2 + D(r)  d\phi^2\,.
\end{equation}
For black hole solutions that support Maxwell-type electromagnetic fields, such as the Reissner--Nordstr\"om family and related spacetimes, the background electromagnetic potential can be written in the general form
\begin{equation}
    V = V_{\mu}\, dx^{\mu} = V_t(r)\, dt \,,
\end{equation}
where \(V_t(r)\) is the electrostatic potential. For the standard Reissner-Nordstr\"om-type geometry, this takes the explicit form
\begin{equation}
    V_t(r) = \frac{Q}{r}\,, \qquad V_r = V_\theta = V_\phi = 0\,,
\end{equation}
with \(Q\) denoting the black hole charge. When a charged test particle of charge \(q\) moves in this background, its interaction with the electromagnetic field is governed by the minimal coupling term
\begin{equation}
    q\, V_{\mu}\, \dot{x}^{\mu} = q\, V_t(r)\, \dot{t}\,,
\end{equation}
which directly enters the effective Lagrangian or Hamiltonian. This form of the potential reflects the spacetime's static, spherically symmetric structure. It plays a crucial role in determining both the electromagnetic contribution to the geometry and the dynamics of charged test particles.

In the equatorial plane of a spherically symmetric spacetime, the motion of charged particles is governed by the Lagrangian
\begin{equation}
\label{lang}
2\mathcal{L} = -f(r)\dot{t}^2 + \frac{1}{h(r)}\dot{r}^2 + D(r)\dot{\phi}^2 - 2qV_t\,\dot{t}\,,
\end{equation}
where an overdot represents differentiation with respect to \(t\), while a prime denotes differentiation with respect to the radial coordinate \(r\).
The dynamical state of the system is fully characterized by the position and its conjugate momentum, defined via the standard relation
\begin{equation}
p_i = \frac{\partial \mathcal{L}}{\partial \dot{\alpha}^i}\,,
\end{equation}
yielding the following non-vanishing components:
\begin{eqnarray}
 p_t&=&-f(r)\dot{t}-qV_t=-E\,,\nonumber\\ p_r&=&\frac{\dot{r}}{h(r)}\,,\label{egn: c.momenta}\\
 p_\phi&=&D(r)\dot{\phi}=L\,. \nonumber
\end{eqnarray}
The constants $E$ and $L$ signify energy and angular momentum, respectively. One can write the Hamiltonian of the system as
\begin{equation}
\label{eqn:ham1}
\mathcal{H}=p_t\dot{t}+p_r\dot{r}+p_\phi\dot{\phi}-\mathcal{L}\,.
\end{equation}
Since the Lagrangian is independent of time, the system admits a conserved Hamiltonian, which can initially be expressed in terms of the generalized velocities as
\begin{equation}
\label{eqn: ham2}
2\mathcal{H} = -E\,\dot{t} + L\,\dot{\phi} + \frac{\dot{r}^2}{h(r)}\,,
\end{equation}
where \(E = -p_t\) and \(L = p_\phi\) are the conserved energy and angular momentum, respectively. Utilizing the expressions for the generalized momenta obtained in Eqs.~(\ref{egn: c.momenta})–(\ref{eqn: ham2})), the Hamiltonian can be recast in momentum space as
\begin{equation}
\label{eqn: hamiltonian}
\mathcal{H} = \frac{-\left(p_t + qV_t\right)^2 + f(r)h(r)\,p_r^2 + f(r)D(r)^{-1}p_\phi^2}{2f(r)}\,.
\end{equation}
The canonical equations of motion derived from the Hamiltonian in Eq.~(\ref{eqn: hamiltonian}) are then given by the standard Hamiltonian flow
\begin{equation}
\frac{dX_i}{d\tau} = \frac{\partial \mathcal{H}}{\partial p_i}\,, \qquad \frac{dp_i}{d\tau} = -\frac{\partial \mathcal{H}}{\partial X_i}\,,
\end{equation}
where \(X_i \in \{t, r, \phi\}\) and \(p_i\) are the associated conjugate momenta, and in this case, the canonical equations of motion reads 
\begin{eqnarray}
\dot{t}&=&\frac{\partial \mathcal{H}}{\partial p_t}=-\frac{p_t+qV_t}{f(r)}\,,\label{tdot}\\
\dot{p}_t &=&-\frac{\partial \mathcal{H}}{\partial t}=0\,,\nonumber\\
\dot{r}&=&\frac{\partial \mathcal{H}}{\partial p_r}=p_rh(r)\,,\label{rdot}\\
\dot{p}_r&=&-\frac{\partial \mathcal{H}}{\partial r}=-\frac{1}{2}\left [p_r^2h(r)^\prime-\frac{2qV_t^\prime(p_t+qV_t)} {f(r)}+\frac{(p_t+qV_t)^2f(r)^\prime}{f(r)^2}-p_\phi^2D(r)^{-2}D(r)^\prime\right]\,,\label{eqn: mots}\\
\dot{\phi}&=&\frac{\partial\mathcal{H}}{\partial p_\phi}=\frac{p_\phi}{D(r)}\,,\nonumber\\ \dot{p}_\phi &=&-\frac{\partial \mathcal{H}}{\partial\phi}=0\,.\nonumber
\end{eqnarray}
The relationship between time and radial coordinate, time and radial momentum are obtained  from the equations of motion (\ref{tdot}) to (\ref{eqn: mots}) as
\begin{eqnarray}
\frac{dr}{dt}&=&\frac{\dot{r}}{\dot{t}}=-\frac{p_rf(r)h(r)}{p_t+qV_t}\,,\nonumber\\
\frac{dp_r}{dt}&=&\frac{\dot{p}_r}{\dot{t}}=-qV_t+\frac{1}{2}\left [\frac{p_r^2f(r)h(r)^\prime}{p_t+qV_t}+\frac{(p_t+qV_t)f(r)^\prime}{f(r)}-\frac{p_\phi^2D^{-2}D^\prime f(r)}{p_t+qV_t}\right]\,. \label{eqn:prdot}
\end{eqnarray}
We now define the dynamical functions \( F_1 = \frac{dr}{dt} \) and \( F_2 = \frac{dp_r}{dt} \). Moreover, the motion of particles in curved spacetime is governed by the normalization condition of the four-velocity, given by \( g_{\mu\nu} \dot{x}^\mu \dot{x}^\nu = \eta \), where \( \eta = 0 \) for massless particles (photons) and \( \eta = -1 \) for massive particles. This normalization imposes a constraint on the metric tensor associated with the line element defined in Eq.~(\ref{ds}), and serves as a key condition in deriving the equations of motion. In this normalisation, we find
\begin{equation}
p_t+qV_t=-\sqrt{f(r)(1+p_r^2h(r)+\frac{p_\phi^2}{D(r)}}\,.
\end{equation}
It immediately implies that equation (\ref{eqn:prdot}) takes a new form as
\begin{eqnarray}
\label{dp}
F_1&=&\frac{p_rf(r)h(r)}{\sqrt{f(r)\left(1+p_r^2h(r)+\frac{p_\phi^2}{D(r)}\right)}}\,,\nonumber\\
F_2&=&-qV_t^\prime-\frac{p_r^2(f(r)h(r))^\prime+f(r)^\prime+p_\phi^2\left(\frac{f(r)}{D(r)}\right)^\prime}{2\sqrt{f(r)\left(1+p_r^2h(r)+\frac{p_\phi^2}{D(r)}\right)}}\,.
\end{eqnarray}
We write the Jacobian $J$ matrix as 
\begin{equation}
\label{eqn:jjc}
J= \begin{pmatrix}
K_{11} & K_{12}\\
K_{21} & K_{22}\\ 
\end{pmatrix}\,,
\end{equation}
with
\begin{equation*}
\begin{aligned}
K_{11} &= \frac{\partial F_1}{\partial r}
  = \frac{p_r \big(f(r) h(r)\big)'}%
         {\sqrt{f(r)\left(1 + p_r^2 h(r) + p_\phi^2 D(r)^{-1}\right)}}
\\
&\quad
  - p_r h(r) f(r)\,
    \frac{f(r)' + p_r \big(f(r) h(r)\big)' + p_\phi^2 \big(D(r)^{-1} f(r)\big)'}%
         {2\left[f(r)\left(1 + p_r^2 h(r) + p_\phi^2 D(r)^{-1}\right)\right]^{3/2}}\,,
\\[10pt]
K_{12} &= \frac{\partial F_1}{\partial p_r}
  = \frac{f(r) h(r)}%
         {\sqrt{f(r)\left(1 + p_r^2 h(r) + p_\phi^2 D(r)^{-1}\right)}}
\\
&\quad
  - \frac{\big(p_r f(r) h(r)\big)^2}%
         {\left[f(r)\left(1 + p_r^2 h(r) + p_\phi^2 D(r)^{-1}\right)\right]^{3/2}}\,,
\\[10pt]
K_{21} &= \frac{\partial F_2}{\partial r}
  = -q\, V_t'' 
    - \frac{f(r)'' + p_r^2 \big(f(r) h(r)\big)'' + p_\phi^2 \big(D(r)^{-1} f(r)\big)''}%
           {2\sqrt{f(r)\left(1 + p_r^2 h(r) + p_\phi^2 D(r)^{-1}\right)}}
\\
&\quad
  + \frac{\big[f(r)' + p_r^2 \big(f(r) h(r)\big)' + p_\phi^2 \big(D(r)^{-1} f(r)\big)'\big]^2}%
         {4\left[f(r)\left(1 + p_r^2 h(r) + p_\phi^2 D(r)^{-1}\right)\right]^{3/2}}\,,
\\[10pt]
K_{22} &= \frac{\partial F_2}{\partial p_r}
  = -\frac{p_r \big(f(r) h(r)\big)'}%
         {\sqrt{f(r)\left(1 + p_r^2 h(r) + p_\phi^2 D(r)^{-1}\right)}}
\\
&\quad
  + p_r h(r) f(r)\,
    \frac{f(r)' + p_r \big(f(r) h(r)\big)' + p_\phi^2 \big(D(r)^{-1} f(r)\big)'}%
         {2\left[f(r)\left(1 + p_r^2 h(r) + p_\phi^2 D(r)^{-1}\right)\right]^{3/2}}\,.
\end{aligned}
\end{equation*}
For a circular orbit, the radius must remain constant, which means that the particle does not have radial motion. Thus, the radial momentum must vanish at the orbital radius as follows
\begin{equation}
\label{prr}
p_r(r_0)=0\,.
\end{equation}
 In addition, the radial force must vanish so that no acceleration drives the particle away from the circular trajectory. Equivalently, the orbital equation for the radial component requires
\begin{equation}
\frac{dp_r}{dt}\bigg|_{r_0}=0\,,
\end{equation}
ensuring that the circular orbit is an equilibrium point of the radial dynamics.  
These two conditions—zero radial momentum and zero radial acceleration—constitute the standard criteria for circular motion in black-hole spacetimes \cite{chandrasekhar1983mathematical, pugliese2011circular}.
Subject to these dynamical constraints, the Lyapunov exponent characterizing the stability of the circular orbit is obtained as the eigenvalue of the Jacobi matrix defined in Eq.~(\ref{eqn:jjc}), and can be written as
\begin{eqnarray}
\label{eqn: Glyapunov}
\lambda^2 &=&
\frac{1}{4} \frac{h(r) \left[ f(r)' + p_\phi^2 \left(D(r)^{-1} f(r)\right)' \right]^2}{f(r) \left(1 + p_\phi^2 D(r)^{-1} \right)^2}\\\nonumber
&-& \frac{1}{2} h(r) \frac{f(r)'' + p_\phi^2 \left(D(r)^{-1} f(r)\right)''}{1 + p_\phi^2 D(r)^{-1}}
- \frac{q V_t'' f(r) h(r)}{\sqrt{f(r)\left(1 + p_\phi^2 D(r)^{-1} \right)}}\,.
\end{eqnarray}
Equation~(\ref{eqn: Glyapunov}) presents the general expression for the Lyapunov exponent, a key diagnostic tool in the analysis of circular orbit stability and the onset of chaotic dynamics for test particles subjected to an external electromagnetic potential \( V_t \) in a spherically symmetric black hole background. This expression accounts for the conserved angular momentum \( p_\phi \) of the particle, as well as the metric functions \( f(r) \), \( h(r) \), and \( D(r) \) that characterise the geometry of spacetime. The expression remains valid irrespective of whether the spacetime satisfies \( f(r) = h(r) \) (as in standard Schwarzschild-type metrics) or \( f(r) \neq h(r) \), thus encompassing a broad class of spherically symmetric black hole solutions.
\subsubsection{Notes on the Jacobi method}\label{nj}
The Jacobi method provides a systematic framework for analysing timelike orbital motion. Within this framework, the radii of circular orbits are obtained from the equilibrium condition (\ref{prr}). For spacetimes whose structure is more involved than Schwarzschild or Reissner–Nordstr\"{o}m, Eq.~(\ref{prr}) may admit multiple and generally transcendental roots. In such cases, one solves Eq.~(\ref{prr}) numerically (for example, with robust root-finding algorithms), imposing the physical requirement:
\(
r_{0} > r_{+},
\)
so that admissible orbits lie outside the event horizon.

The conserved quantities associated with circular motion (energy \(E\) and angular momentum \(L\)) must be real and finite. Importantly, their admissible values are fixed consistently by the spacetime parameters rather than being independent free parameters. Therefore, when employing the Jacobi method, it is significantly crucial to explicitly derive the conserved quantities corresponding to each permissible value of $r_{0}$. This ensures that the stability analysis, represented by the Lyapunov exponent, is performed solely for physically valid orbital trajectories. To avoid ambiguity, one must therefore compute the conserved quantities associated with each admissible root explicitly.

\subsubsection{Physical constraints on circular-orbits}\label{prop:consistency}
Let \(r_{0}\) denote the radius of a circular orbit determined by the orbital conditions of the spacetime. 
This radius is fixed uniquely by the geometric parameters \(\{\xi_i\}\), such as the black hole mass, charge, cosmological constant, and any additional matter fields, as follows
\begin{equation}
r_{0} = r_{0}(\xi_{1},\dots,\xi_{n}) \, .
\end{equation}
The associated angular momentum is constrained by the same parameters and may be expressed equivalently as a function of the orbital radius in the following way 
\begin{equation}
L_{0} = L(r_{0};\,\xi_{1},\dots,\xi_{n}) \, .
\end{equation}
The circular-orbit conditions form a coupled system in \(r\) and \(L\), whose solutions determine the admissible pairs \((r_{0}, L_{0})\).
Each real solution for \(r_{0}\) selects a unique value of the angular momentum \(L_{0}\), which must be computed explicitly.
Physical circular orbits are identified by the requirements \cite{pradhan2011circular, pugliese2011circular}
\begin{equation}
r_{0} \in \mathbb{R},
\qquad
r_{0} > r_{+},
\label{r00}
\end{equation}
ensuring that the orbit lies outside the event horizon.

Any subsequent analysis, including stability criteria and the evaluation of Lyapunov exponents, must therefore be formulated in terms of the consistent pairs \((r_{0}, L_{0})\).
In particular, the Lyapunov exponent may be written schematically as
$\lambda = \lambda(L_{0},\, r_{0}) \,.$ By treating $r_0$ and $L_0$ independently can produce configurations that are not physically admissible within the probe/orbital assumptions of the setup, and can therefore lead to misleading conclusions. For this reason, the explicit mapping \(r_{0} \mapsto L_{0}\) is adopted throughout this work.
\subsubsection{Exact formula for angular momentum of charged particles at equilibrium orbits}
In Sec.~\ref{sec2a}, we explained why equilibrium circular orbits cannot have a radial momentum component, as required by the circular orbit condition (\ref{prr}). Under this condition, one can construct a quadratic equation for the angular momentum at equilibrium orbits by defining $p_\phi^2 = L_0^2 = X_1$, we get
\begin{equation} \label{quadr}
\alpha_1 X_1^2 + \beta_1 X_1 + \gamma_1 = 0\,,
\end{equation}
with the coefficients given by
\begin{eqnarray}
\alpha_1 &=& \left[\left(\frac{f(r)}{D(r)}\right)'\right]^2\,, \label{coeff:alpha} \\[1ex]
\beta_1  &=& 2\left(\frac{f(r)}{D(r)}\right)' f'(r)
           - \frac{4 q^2 V_t'(r)^2 f(r)}{D(r)}\,, \label{coeff:beta} \\[1ex]
\gamma_1 &=& f'(r)^2 - 4 q^2 V_t'(r)^2 f(r)\,. \label{coeff:gamma}
\end{eqnarray}
The solution for the angular momentum is therefore given by
\begin{equation}\label{eq:L2-general}
L_0^2 \;=\; \frac{-\beta_1 \pm \sqrt{\beta_1^2 - 4 \alpha_1 \gamma_1}}{2 \alpha_1}\,.
\end{equation}

Equation (\ref{eq:L2-general}) provides an exact expression for the angular momentum of orbiting timelike particles in equilibrium. In general, determining the exact location of circular timelike orbits is a non-trivial task, particularly within the Jacobi framework. Nevertheless, their existence can be inferred from the underlying physical consistency requirements.
In particular, physical admissibility demands that the angular momentum be real and finite. The finiteness condition requires that
\(
    \alpha_1 > 0,
\)
while the requirement of reality imposes
\(
    \beta_1^2 - 4\alpha_1\gamma_1 > 0\,,~
    -\beta_1 \pm \sqrt{\beta_1^2 - 4\alpha_1\gamma_1} > 0.
\)
If circular timelike orbits exist, they must satisfy all of these constraints simultaneously. We note that the first condition determines the possible radii of circular motion (as will be shown later), whereas the latter condition serves to constrain additional parameters that are not otherwise fixed by the intrinsic spacetime geometry. Let the radius satisfying \(\alpha_1(r_0) > 0\) be denoted by \(r_{0+}\). The physical circular orbit can then be parameterized as
\begin{equation}
r_{0} = r_{0+} + \epsilon\,,
\end{equation}
where \(\epsilon \ge 0\) is a small adjustable parameter that controls the orbit’s proximity to the event horizon. This prescription provides a convenient and physically transparent way to parametrize timelike circular orbits both near and far from the horizon, while simultaneously ensuring that the condition for physical consistency is strictly satisfied, that the angular momentum remains finite, and that the orbit lies within the admissible region of the spacetime. For massless particles, the condition for the photon sphere is \(\alpha_1 = 0\), corresponding to \(\epsilon = 0\). In particular, for a massive particle to remain at \(\alpha_1=0\), it would need to travel at the speed of light, which is physically unattainable. As a result, only massless particles can occupy the photon sphere, whereas timelike circular orbits necessarily satisfy \(\alpha_1 > 0\). We note that the criterion for circular orbits (\(\alpha_{1} > 0\)) coincides with that for homoclinic trajectories, as will become evident in the subsequent Lyapunov exponent analysis. Hence, all circular orbits with physically realistic angular momentum are homoclinic, which also agrees with Ref.~\cite{hashimoto2017universality, jeong2023homoclinic}.

\subsubsection*{Basic examples}

\begin{itemize}
    \item \textbf{Schwarzschild black hole.}  
    The metric functions are 
    \[
    f(r) = 1 - \frac{2m}{r}\,, \qquad D(r) = r^{2}\,.
    \]
   In the absence of Maxwell fields (i.e., \(Q = q = 0\)), and since our analysis is restricted to circular orbits, we set \(\theta = \pi/2\) throughout. In this case,
\[
\alpha_{1} = \frac{4 (r_{0} - 3m)^{2}}{r_{0}^{8}}\,,
\]
and the condition \(\alpha_{1} > 0\) leads to the nontrivial constraint \(r_{0} > 3m\). The corresponding angular momentum is
 \[
    L_0 = \sqrt{\frac{m r_{0}^{2}}{r_{0} - 3m}}.
    \]

    \item \textbf{Reissner--Nordstr\"om black hole.}  
    The metric functions are
    \[
    f(r) = 1 - \frac{2m}{r} + \frac{Q^{2}}{r^{2}}, \qquad D(r) = r^{2}\,.
    \]
    Here,
    \[
    \alpha_{1} = \frac{4 \big(r_{0}(r_{0} - 3m) + 2Q^{2}\big)^{2}}{r_{0}^{10}}\,,
    \]
    and the angular momentum is given by
    \[
    L_0 =
    \sqrt{
    \frac{
    r_{0}^{2}\!\Big(
    2m r_{0}^{3} + r_{0}^{2}\big[(q^{2} - 2)Q^{2} - 6m^{2}\big]
    - 2mQ^{2}(q^{2} - 5)r_{0} + (q^{2} - 4)Q^{4}
    + \beta_{2}
    \Big)
    }{
    2\big[r_{0}(r_{0} - 3m) + 2Q^{2}\big]^{2}
    }
    }\,,
    \]
    where
    \[
    \beta_{2}
    = \sqrt{\,q^{2}Q^{2}\big[r_{0}(r_{0} - 2m) + Q^{2}\big]^{2}
    \big[4r_{0}(r_{0} - 3m) + (q^{2} + 8)Q^{2}\big]}\,.
    \]
    By using the condition $\alpha_{1} > 0$, the timelike circular orbits are obtained as
    \[
    r_{0\pm} > \frac{3m \pm \sqrt{9m^{2} - 8Q^{2}}}{2}\,,
    \]
    \end{itemize}
which are the well-known Reissner-Nordstr\"om results~\cite{mokdad2017reissner, lambiase2024weak}. 
These cases confirm that Eq.~(\ref{eq:L2-general}) not only recovers the expected classical behavior but also extends naturally to more intricate spacetimes with electromagnetic or matter fields.
\subsection{Lyapunov exponents from the effective potential of charged particles}  
We start by deriving the effective potential that characterizes particle motion near the black hole, following the standard approach presented in Refs.~\cite{cardoso2009geodesic, lei2021chaos, kumara2024lyapunov, jeong2023homoclinic}. In this section, we build on this well-established procedure and extend it to a more general case in which the orbiting test particle carries an electric charge. For a spacetime with signature $(-,+,+,+)$, a static and spherically symmetric metric maintains the form given in Eq.~(\ref{ds}), with the same Lagrangian equation of motion as in Eq.~(\ref{lang}). Similarly, by using the normalization condition for the four-velocity vectors, as mentioned earlier, together with Eqs.~(\ref{egn: c.momenta}) and (\ref{eqn:ham1}), we express the radial motion of the charged particle in terms of the four-velocity $\dot{x}^\mu = \frac{dx^\mu}{d\tau}$, with the radial component
\begin{equation}
\dot{r}^2 + V_{\rm eff} = 0\,,
\end{equation}
where the effective potential is given by
\begin{equation}\label{eq:rdot_isolated}
V_{\rm eff} = h(r)\left[\eta + \frac{\widetilde E^{2}}{f(r)} - \frac{L^{2}}{D(r)}\right]\,,
\end{equation}
with $\widetilde E= E - q V_{t}(r)$. We shall therefore focus on the special case $\eta=-1,$ for massive particles in this derivation. We shall also explain later on the conditions under which this derivation can apply to massless particles as well.

For circular orbits, the effective potential must possess a local maximum, which is equivalent to requiring that both the radial velocity and radial acceleration vanish as follows: 
\begin{equation}
\label{trt}
\dot{r}=0\,, \qquad \ddot{r}=0\,.
\end{equation}
These conditions ensure that the radial force is zero, keeping the particle momentarily at a fixed radius. A convenient and physically transparent way to determine both the circular-orbit radius and its associated angular momentum was used by Ref.~\cite{hashimoto2023causality}, namely, by locating the value of the angular momentum for which the effective potential develops a local maximum.
In the following, we generalize this method to show that both the circular-orbit radius and the corresponding angular momentum, previously derived in Eq.~(\ref{eq:L2-general}), arise naturally from this procedure. We shall later verify that this approach reproduces the angular-momentum relations familiar from particle dynamics.

By introducing the definition \(X_1 = L_0^{2}\), and imposing the circular–orbit condition (\ref{trt}) on Eq.~\eqref{eq:rdot_isolated}, we have
\begin{equation}\label{eq:Et}
\widetilde E^{2} = f(r)\!\left(\frac{X_1}{D(r)} + 1\right)\,.
\end{equation}
Additionally, by using Eq.~(\ref{trt}) once again, we obtain
\begin{equation}\label{eq:d}
2\widetilde E\,\widetilde E'\,f(r) - \widetilde E^{2} f'(r) + X_1\frac{f(r)^{2} D'(r)}{D(r)^{2}} = 0\,,
\end{equation}
and by substituting \eqref{eq:Et} into \eqref{eq:d}, we get
\begin{align}
2\widetilde E\,\widetilde E'\,f(r) 
 - f(r)\Big(\frac{X_1}{D(r)}+1\Big)f'(r) 
 + X_1\frac{f(r)^{2} D'(r)}{D(r)^{2}} &= 0\,.
 \label{eq:sub1}
\end{align}
By rearranging Eq.~\eqref{eq:sub1} in such a way that all terms proportional to \(X_1\) appear on the left-hand side, and using \(\widetilde E'=-qV_t'\), we arrive at
\begin{equation}
\label{eq:X}
X_1 
= \frac{D(r)^{2}\,\big[\,2 q\,\widetilde E(r)\,V_t'(r) + f'(r)\,\big]}
       {\,f(r) D'(r) - f'(r) D(r)\,}\;.
\end{equation}

For neutral particles (\(q=V_t=0\)), this expression reduces to the familiar results reported in Refs.~\cite{touati2024lyapunov, kumara2024lyapunov, hashimoto2017universality, lei2022circular, ali2025evaluation}. 
This demonstrates explicitly that the effective potential and Jacobi methods are equivalent; the choice between them is one of convenience, as both approaches impose the same constraints on the angular momentum.

When \(q\neq 0\) and \(V_t\neq 0\), the essential point is that the particle’s charge couples to the black hole’s electromagnetic field, thereby modifying the angular momentum directly; this, in turn, shifts the location of circular orbits indirectly through the dependence of the motion on \(X_1\). 
Notably, substituting Eq.~(\ref{eq:Et}) into Eq.~(\ref{eq:X}) yields exactly the quadratic Eq.~(\ref{quadr}) in \(X_1\), whose corresponding solution is given in Eq.~(\ref{eq:L2-general}).
Additionally, we note that the physically admissible condition in this case (also discussed in Ref.~\cite{javed2025joule}), is expressed as
\begin{equation}
    f(r) D'(r) - f'(r) D(r) > 0\,,
\end{equation}
which we shall later denote by
\begin{equation}
\label{eq:condition}
D_1(r) = f(r) D'(r) - f'(r) D(r) > 0\,.
\end{equation}
This is equivalent to the condition \(\alpha_1 > 0\) and can be used to identify timelike circular orbits. For null circular orbits, the standard and widely used photon sphere condition is expressed as
\begin{equation}
\label{eqn:circ}
f(r) D'(r) - f'(r) D(r) = 0\,.
\end{equation}
 It is important to clarify that Eq.~\eqref{eq:X} is obtained under the timelike normalization and therefore applies strictly to massive circular orbits. In this expression the limit \(f(r) D'(r) - f'(r) D(r) \to 0\) causes \(L_0^2\) to diverge. This divergence should not be interpreted as a photon possessing infinite angular momentum; instead, it reflects the fact that the timelike normalization breaks down when approaching a null orbit. For massless particles, the overall scale of the conserved quantities \(E\) and \(L\)
is arbitrary, and the physically meaningful quantity is the impact parameter \cite{ma2020bounds, shaikh2019strong} given by
$b = \frac{L}{E} = \frac{r_0}{\sqrt{f(r_0)}}\,,$ which remains finite at the photon-sphere radius. 

The stability of circular orbits can be assessed by evaluating the Lyapunov exponent. An orbit is stable if its proper-time Lyapunov exponent $(\lambda_{\rm p})$ or its coordinate-time counterpart $(\lambda_{\rm c})$ is imaginary; otherwise, it is unstable~\citep{touati2024lyapunov,kumara2024lyapunov}. In general, instability is characterized by the condition  
\begin{equation}
\label{lcon}
    \frac{\lambda_{\rm p}}{\lambda_{\rm c}}>0\,.
\end{equation}
The Lyapunov exponents can be expressed in two equivalent forms. The proper-time Lyapunov exponent, $\lambda_{\rm p}$, and the coordinate-time Lyapunov exponent, $\lambda_{\rm c}$, are given by \cite{touati2024lyapunov, kumara2024lyapunov}
\begin{equation}
\label{ll1}
\lambda_{\rm p} =\sqrt{\frac{-V_{\rm eff}^{\prime\prime}}{2}}\,, 
\end{equation}
and 
\begin{equation}
\label{eqn:Lyap}
\lambda_{\rm c} = \sqrt{\frac{-V_{\rm eff}^{\prime\prime}}{2\dot{t}^2}}\,,
\end{equation}
respectively. Equations~(\ref{ll1}) and~(\ref{eqn:Lyap}) provide the basis for analyzing the stability of circular motion around black hole spacetimes. Since the MSS bound relies on boundary (coordinate) time, any instability in proper time is not interpreted as a violation of the chaos-bound. In particular, they allow one to determine whether small perturbations around equilibrium orbits grow or decay, thereby providing a precise tool for studying both the stability and the onset of chaotic dynamics in black hole spacetimes. 

\section{Chaotic dynamics and stability analysis in black hole spacetimes}
\label{san}
After developing a consistent framework in the previous section, based on both effective potential and Jacobian matrix approaches for determining the Lyapunov exponent, we now employ it to two intricate black hole spacetimes to better understand our formalism. This section serves two main objectives: first, to demonstrate that the proposed approach eliminates apparent chaos-bound violation due to unconstrained orbital parameters; and second, to highlight cases where real violations arise from the curvature corrections inherent to extensions of GR. For this purpose, we investigate two models that incorporate both classical and extensions of GR effects: the charged Kiselev black hole surrounded by a string cloud and quintessence, and the charged $f(R)$ black hole in the presence of a cosmological constant. By considering these cases, we perform a comprehensive analysis of the reliability of our formalism and its physical significance for describing the chaotic dynamics in gravitational systems.
\subsection{chaos-bound in charged Kiselev black hole with a cosmological constant}
This section focuses on the Kiselev spacetime, which is simultaneously surrounded by a string cloud and a cosmological constant, and we consider it here as a case study. This choice is motivated by the fact that such a configuration has not been systematically investigated in the literature. Furthermore, by appropriately tuning the black hole parameters, one recovers several well-known black hole solutions, some of which have already been analyzed in the context of chaos-bound violations~\cite{gao2022chaos, li2025chaotic}, thereby enabling a meaningful comparative study. The more general case, where the line element is extended to incorporate both string cloud and cosmological constant contributions, has been examined from a thermodynamic perspective in Ref.~\cite{zafar2025thermodynamic}, which also provides additional details about the underlying geometry. For completeness, following Refs.~\citep{toledo2019some, liang2021remarks}, we recall the general metric ansatz given in Eq.~(\ref{ds}), with the corresponding metric components expressed as 
\begin{equation}
\label{lne}
f(r)=h(r)= 1 - a - \frac{2m}{r} + \frac{Q^2}{r^2} - \frac{\Lambda r^2}{3} - \frac{\alpha}{r^{3\omega + 1}}\,, 
\quad B(r)=r^2\,, 
\quad D(r)=r^2\sin^2\theta\,,
\end{equation}
where \(m\) is the black hole mass, \(Q \geq 0\) the electric charge, and \(\Lambda\) the cosmological constant. The parameter \(a\) measures the contribution of the surrounding string cloud, while \(\alpha\) is a normalization constant associated with the energy density of the anisotropic fluid characterized by the equation-of-state parameter \(\omega\).  

The matter component satisfies the relation \(p = \omega \rho\), where \(\omega \in \mathbb{R}\) may take values such as \(\omega=0\) (dust), \(\omega=-1\) (cosmological constant), or \(\omega=-2/3\) (quintessence). Following Kiselev~\cite{kiselev2003quintessence}, the energy density of the anisotropic fluid takes the form
\begin{equation}
   \rho(r) = -\frac{3\alpha \,\omega}{2\, r^{3(1+\omega)}}\,.
\end{equation}
In this study, we adopt the convention in which $\omega < 0$ and $\alpha > 0$, ensuring that the weak energy condition $\rho(r) \geq 0$ is satisfied and the energy density remains positive definite. 
The parameter \(a\) must satisfy  
\(
0 \leq a <1
\)\,.
The parameter $a$ cannot approach unity if one wishes to preserve the correct spacetime signature and a reasonable asymptotic structure. While $a=0$ corresponds to the absence of strings, values of $a$ approaching unity tend to drive the metric towards a degenerate configuration, as $f(r) \to 0$ at spatial infinity, suggesting a potentially pathological behavior.

The specific case of interest in this study corresponds to the choice \(\omega = -\tfrac{2}{3}\). For this value of the state parameter, the metric function simplifies to
\begin{equation}
\label{fk}
f(r) = 1 - a - \frac{2m}{r} + \frac{Q^2}{r^2} - \frac{\Lambda r^2}{3} - \alpha r\,,
\end{equation}
and the surface gravity at the outer horizon is determined by
\begin{equation}
    \kappa = \frac{1}{2}\, f(r)^\prime \Big|_{r_+} \,.
\end{equation}
This black hole solution represents a Reissner--Nordstr\"om black hole surrounded simultaneously by a cloud of strings and a quintessence field with a cosmological constant. This special case defines the specific Kiselev black hole background that will be the focus of our analysis. Notably, by assigning specific values to the parameters in Eq.~(\ref{lne}), one can recover several distinct geometries. For example, setting \(a = \Lambda = 0\) yields the Reissner--Nordstr\"om black hole surrounded by a quintessence field, a configuration previously reported to exhibit chaos-bound violations~\cite{gao2022chaos}. Other special cases will be discussed later.

\subsubsection{Jacobi method for the Reissner-Nordstr\"om black hole surrounded by quintessence} \label{jjjj}
The Lyapunov exponent is computed from the Jacobi-matrix method from Eq.~(\ref{eqn: Glyapunov}) as shown in Appendix~\ref{app:A}.
The angular momentum expressions obtained via Eq.~(\ref{eq:L2-general}) determine the location of circular orbits. As will be discussed later, the denominator appearing here is consistent with the conditions for homoclinic orbits. To ensure that $L_0$ remains real and finite, the denominator must satisfy the inequality $D_1(r_0) > 0\,$. In the limit $\alpha \to 0$, this condition reduces to the familiar Reissner--Nordstr\"om case, expressed as  
\begin{equation}
    r_0>\frac{3m\pm\sqrt{9m^2-8Q^2}}{2}\,.
\end{equation}
In the special case where the normalization constant $\alpha \neq 0$, the condition becomes more involved. For completeness, we state it here as  
\begin{equation}
\label{e30}
    r_0>\frac{2}{3 \alpha }+\frac{D_2}{3 \,2^\frac{1}{3} \alpha }-\frac{{2^\frac{1}{3}} (18 \alpha  m-4)}{3 \alpha  D_2}\,,
\end{equation}
with $$D_2=\left[-108 \alpha  m+\sqrt{4 (18 \alpha  m-4)^3+\left(-108 \alpha  m+108 \alpha ^2 Q^2+16\right)^2}+108 \alpha ^2 Q^2+16\right]^\frac{1}{3}.$$
The event horizon of the black hole is independent of the charge carried by the orbiting test particle. Although the particle may be charged, its interaction with the background field affects only dynamical quantities and does not alter the location of the circular orbits. In particular, the angular momentum depends explicitly on the test charge (see Eq.~\ref{ang1} in Appendix~\ref{app:A}). For this dependence to remain physically meaningful, the test charge must satisfy the inequality
\begin{equation}
\label{cc}
     q \ll Q < m\,,
\end{equation} 
which follows from the physical admissibility, ensuring that the angular momentum remains real. This constraint is consistent with the results discussed in Ref.~\cite{hashimoto2017universality}, which demonstrated that, for regular electromagnetic or scalar forces, the maximal Lyapunov exponent is insensitive to both the external force and the particle mass, thereby leading to universal saturation of the chaos-bound.
\begin{figure}[htb]
    \centering
    \includegraphics[width=0.45\textwidth]{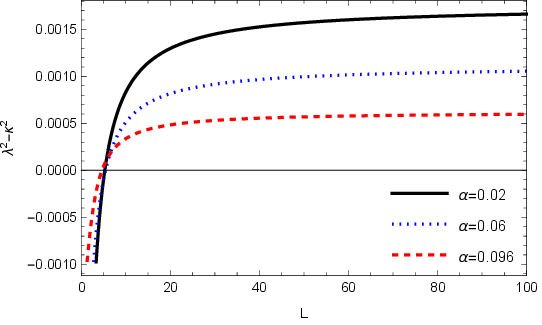}
    \caption{Apparent violations of the chaos-bound for fixed parameters \(m = 2\), \(q = 15\), and \(Q = 1.95\). Here, we present different curves corresponding to different values of the normalization constant $\alpha$, such as for $\alpha=0.02$ (solid black curve), $\alpha=0.06$ (dotted blue curve), and $\alpha=0.096$ (dashed red curve). Unrealistically high angular momentum leads to orbits very close to the horizon, causing apparent violations.  However, these apparent violations disappear when the angular momentum is determined self-consistently from the circular orbit condition as given in Table \ref{tab:tjac}. 
}
    \label{fig:apcb}
\end{figure}

\begin{table*}[h]
\caption{\label{tab:tjac}Horizon dynamics and chaos-bound saturation for a Kiselev spacetime surrounded by quintessence, using the Jacobi matrix method. Here $m=1~Q=0.5,~q=9\times 10^{-22},~\epsilon=0.0005$.}

\begin{ruledtabular}
\begin{tabular}{ccccccc}
$\alpha$ & $r_{+}$ & $r_{0}$ & $L_0$ & $|\kappa|$ & $\lambda$ & $\lambda^{2}-\kappa^{2}$ \\
\hline
0.5 & 0.133291 & 0.176604 & 1.32117 & 49.53320 & 33.7505 & -1314.43788 \\
1.0 & 0.132629 & 0.176093 & 1.32788 & 50.80940 & 34.5723 & -1386.35726 \\
1.5 & 0.131986 & 0.175591 & 1.33412 & 52.07800 & 35.3894 & -1459.70474 \\
2.0 & 0.131361 & 0.175097 & 1.33993 & 53.33910 & 36.2021 & -1534.46627 \\
2.5 & 0.130754 & 0.174611 & 1.34534 & 54.59320 & 37.0106 & -1610.62852 \\
\end{tabular}
\end{ruledtabular}
\end{table*}
Table~\ref{tab:tjac} summarizes the analysis of circular timelike (coordinate) orbits obtained using the Jacobi method. The results indicate that, once the circularity conditions and associated dynamical constraints are imposed consistently, no violation of the chaos bound is observed within the explored parameter range. In this constrained framework, the angular momentum is not treated as an independent input but is determined by the circular-orbit conditions, which naturally restrict it to the physically admissible domain. This self-consistent prescription ensures that the Lyapunov exponent is evaluated on equilibrium configurations compatible with the underlying spacetime geometry.

A subtle but important point concerns the prescription used to determine the circular orbit and its associated conserved quantities. If the orbital radius and angular momentum are not determined simultaneously from the full set of circularity conditions, different parameter assignments may lead to qualitatively different conclusions regarding the chaos bound. 

This feature is illustrated in Fig.~\ref{fig:apcb}. For the fixed parameters \(m = 2\), \(q = 15\), and \(Q = 1.95\), with \(\alpha = 0.02,\, 0.06,\) and \(0.096\), one obtains the pairs \((r_{+}, r_{0}) = (1.3886,\,1.4099)\), \((1.5768,\,1.5998)\), and \((1.8255,\,1.8486)\), respectively, where the radii \(r_{0}\) are determined numerically from Eq.~\eqref{prr}. If, in this procedure, a fixed numerical value \(L = 5\) is adopted to compute the circular orbit, the resulting configurations place the orbit very close to the event horizon. In this case, the quantity \(\lambda^{2}(r_{0}) - \kappa^{2}\) can exhibit behavior suggestive of a chaos-bound violation, differing from the expectations of Ref.~\cite{hashimoto2017universality}. 
However, when the angular momentum is instead determined self-consistently from the circular-orbit constraints, the allowed values of \(L\) differ from the fixed choice \(L=5\), and the corresponding behavior of \(\lambda^{2}(r_{0}) - \kappa^{2}\) is modified accordingly. In particular, Table~\ref{tab:tjac} shows that the angular momentum does not attain such large values for this black hole configuration, and the choice \(q = 15\) does not yield physically admissible circular trajectories within the constrained framework.

The distinction therefore does not concern the relevance of earlier circular-orbit analyses, but rather the parameter prescription used in implementing the circularity conditions. In the present work, both the orbital radius and the associated conserved quantities are determined simultaneously and consistently from the same dynamical conditions. This ensures that the Lyapunov exponent is evaluated on a fully self-consistent circular equilibrium configuration, thereby isolating instability effects that arise intrinsically from the spacetime geometry.

For clarity, we note that when the circular radius \(r_{0}\) is obtained under a specific assumption for \(L\), any subsequent analysis of \(\lambda^{2}(r_{0}) - \kappa^{2}\) must employ the same value of \(L\) to maintain dynamical consistency. Treating these quantities within a unified prescription eliminates potential ambiguities and provides a transparent framework for assessing the validity of the chaos bound.

In the following section, we conduct a systematic study of the chaos bound for black holes in Anti-de Sitter (AdS) backgrounds, including cases with an additional string cloud. All parameters are treated self-consistently and are fully constrained by the spacetime geometry. We further examine whether genuine chaos-bound violations can arise in de Sitter (dS) backgrounds. Because the angular-momentum condition uniquely fixes the test charge such that it does not introduce an independent degree of freedom, as we have shown in Eq.~(\ref{cc}), it is omitted from the subsequent analysis. Moreover, before proceeding to the fully constrained analysis, we emphasize that our approach may be viewed as imposing additional consistency conditions on circular geodesics. Earlier studies may in some cases be interpreted as evaluating a local instability at radial turning points and, in that sense, remain physically meaningful. The purpose of the present framework is therefore not to discard such analyses but to clarify the parameter regimes in which genuine circular-orbit instabilities occur.
\subsubsection{Massive particles in Kiselev spacetime surrounded by strings, quintessence, and a cosmological constant}  
\label{sectim}
The locations of circular orbits can be determined in two equivalent ways: by imposing the instability conditions encoded in the Lyapunov exponent, or by enforcing the requirement of physically admissible angular momentum. In both approaches, the analysis is governed by the same function \(D_{1}(r_{0})\). For timelike circular orbits, we further require that \(D_{1}(r_{0}) > 0\).  The effective potential governing the dynamics of massive particles, derived from using Eqs.~(\ref{eq:rdot_isolated}) and (\ref{fk}) (see more details in Appendix~\ref{app:B}), paving the way for the derivation of the corresponding Lyapunov exponents and angular momentum.

From the perspective of the Lyapunov exponent, particularly for timelike orbits, unstable circular orbits exist whenever 
\(
D_{1}(r_{0}) > 0.
\)
In the case of the Reissner--Nordstr\"om black hole surrounded by a string cloud, this inequality reduces to

\begin{equation}
        r_0> \frac{-3m\pm\sqrt{8 a Q^2+9 m^2-8 Q^2}}{2 (a-1)}.
\end{equation}
For the Reissner--Nordstr\"om black hole surrounded by quintessence without strings, one recovers Eq.~(\ref{e30}), which was derived using the Jacobi matrix method. In the presence of a string cloud, however, the corresponding condition acquires a slightly more intricate form, given by
\begin{equation}
    r_0>\frac{1}{6 \alpha }\left[\frac{2^{5/3} \left(2 a^2-4 a-9 \alpha  m+2\right)}{(\text{G1}+\text{G2})^\frac{1}{3}}-4 a+2 [2(\text{G1}+\text{G2})]^\frac{1}{3}+4\right]\,,
\end{equation}
where 
\begin{equation*}
    \text{G2}=-4 a^3+12 a^2+27 \alpha  a m-12 a-27 \alpha  m+27 \alpha ^2 Q^2+4\,,
    \end{equation*}
    and 
    \begin{equation*}
        \text{G1}= \frac{\sqrt{4 \left(-4 a^3+12 a^2+3 a (9 \alpha  m-4)-27 \alpha  m+27 \alpha ^2 Q^2+4\right)^2
        +\left(18 \alpha  m-4 (a-1)^2\right)^3}}{2}\,.
\end{equation*}

\begin{table*}[ht]
\centering
\caption{\label{tab:tcase1} Circular orbits and bounds on the timelike Lyapunov exponents for black holes surrounded by quintessence without strings. Parameters: $m=2,~a=0,~Q=0.5~\epsilon=0.01$ and $\Lambda=0$.}
\begin{ruledtabular}
\begin{tabular}{cccccccc}
$\alpha$ & $r_+$ & $r_0$ & $L_0$ & $\lambda_{\rm p}$ & $\lambda_{\rm c}$ & $|\kappa|$ & $\lambda_{\rm c}^2 - \kappa^2$ \\
\hline
0.20 & 0.063495 & 0.084514 & 0.096778 & 286.723 & 299.156 & 480.628 & $-178491$ \\
0.60 & 0.063468 & 0.084493 & 0.096882 & 287.323 & 229.570 & 481.634 & $-179269$ \\
0.96 & 0.063445 & 0.084474 & 0.096976 & 287.864 & 229.943 & 482.539 & $-179971$ \\
\end{tabular}
\end{ruledtabular}
\label{tab:lambda_results}
\end{table*}
\begin{table*}[ht]
\centering
\caption{\label{tab:tcase3}Circular orbits and bounds on the timelike Lyapunov exponents for stringy black holes surrounded by quintessence. $m=2,~Q=0.5,~\alpha=0.20,~\epsilon=0.01$ and $\Lambda=1$.}
\begin{ruledtabular}
\begin{tabular}{cccccccc}
$a$ & $r_+$ & $r_0$ & $L_0$ & $\lambda_{\rm p}$ & $\lambda_{\rm c}$ & $|\kappa|$ & $\lambda_{\rm c}^2 - \kappa^2$ \\
\hline
0.00 & 0.063494 & 0.084514 & 0.096797 & 286.755 & 229.170 & 480.683 & $-178538$ \\
0.01 & 0.063483 & 0.084502 & 0.096788 & 286.943 & 229.309 & 470.163 & $-168470$ \\
0.10 & 0.063390 & 0.084392 & 0.099335 & 288.649 & 234.120 & 483.869 & $-179317$ \\
0.50 & 0.062982 & 0.083910 & 0.096382 & 296.242 & 236.244 & 496.596 & $-190796$ \\
\end{tabular}
\end{ruledtabular}
\end{table*}

\begin{table}[ht]
\centering
\caption{\label{tab:tcase4} Circular orbits and bounds on the timelike Lyapunov exponents for stringy black holes surrounded by quintessence with parameters $m=2,~Q=0.5,~\alpha=0.20,~\epsilon=0.01$ and $\Lambda=0$.}
\begin{ruledtabular}
\begin{tabular}{cccccccc}
$a$ & $r_+$ & $r_0$ & $L_0$ & $\lambda_{\rm p}$ & $\lambda_{\rm c}$ & $|\kappa|$ & $\lambda_{\rm c}^2 - \kappa^2$ \\
\hline
0.00 & 0.063495 & 0.084514 & 0.096778 & 286.723 & 229.153 & 480.628 & $-178491$ \\
0.01 & 0.063485 & 0.084502 & 0.096769 & 286.911 & 229.297 & 480.946 & $-178732$ \\
0.10 & 0.063391 & 0.084392 & 0.096695 & 288.617 & 230.569 & 483.804 & $-180904$ \\
0.50 & 0.062983 & 0.083910 & 0.093618 & 296.211 & 236.228 & 496.533 & $-190741$ \\
\end{tabular}
\end{ruledtabular}
\end{table}

\subsubsection{Massless particles for null geodesics}
Since there is no proper time for null geodesics, the attention here has been shifted to the coordinate-time domain. 
The location of the null-like circular orbits (photon sphere) is obtained exactly from the condition given in Eq.~(\ref{eqn:circ}) as
\begin{equation}
r_0 = \frac{2^{2/3}}{\alpha} \cdot \frac{ 
-3 \alpha m + \dfrac{2}{3}(a - 1)^2 + \left[ -2a + \left( \mathcal{A} + \mathcal{B} \right)^{1/3} + 2 \right] \cdot \dfrac{1}{2} \left( \mathcal{A} + \mathcal{B} \right)^{1/3}
}{
\left( \mathcal{A} + \mathcal{B} \right)^{1/3}
}\,,
\end{equation}
where
\[
\mathcal{A} = 54 Q^2 \alpha^2 - 8a^3 + 24a^2 + 54 a \alpha m - 24a - 54 \alpha m + 8\,,
\]
and
\[
\mathcal{B} = 2 \sqrt{2 \left[ 9 \alpha m - 2 (a - 1)^2 \right]^3 + \left( 27 Q^2 \alpha^2 - 4a^3 + 12a^2 + 27 a \alpha m - 12a - 27 \alpha m + 4 \right)^2}\,.
\]
For the generic Reissner-Nordstrom black hole surrounded by a cloud of strings, the photon sphere is expressed as 
\begin{equation}
    r_0= \frac{-3m+\sqrt{8 a Q^2+9 m^2-8 Q^2}}{2 (a-1)}\,.
\end{equation}
\begin{table*}[ht]
\centering
\caption{\label{tab:case1} Circular geodesics of massless particles in a generic Kiselev spacetime surrounded by quintessence, in the absence of a cosmological constant and string cloud. The parameters are $\Lambda=0,~a=0,~Q=0.5$, and $m=2$.}\begin{ruledtabular}
\begin{tabular}{cccccc}
$\alpha$ & $r_+$ & $r_0$ & $|\kappa|$ & $\lambda_{\rm c}$ & $\lambda_{\rm c}^2 - \kappa^2$ \\
\hline
0.20 & 0.0634951 & 0.084514 & 480.628 & 331.103 & $-121374$ \\
0.60 & 0.0634687 & 0.084493 & 481.634 & 331.760 & $-121907$ \\
0.96 & 0.0634450 & 0.084474 & 482.539 & 332.350 & $-122387$ \\
\end{tabular}
\end{ruledtabular}
\end{table*}

\begin{table*}[ht]
\centering
\caption{\label{tab:case3} Null circular geodesics of a charged black hole in a Kiselev spacetime surrounded by a string cloud and quintessence. The parameters are $m=2,~Q=0.5,~\alpha=0.20$, and $\Lambda=0$.}
\begin{ruledtabular}
\begin{tabular}{cccccc}
$a$ & $r_+$ & $r_0$ & $|\kappa|$ & $\lambda_{\rm c}$ & $\lambda_{\rm c}^2 - \kappa^2$ \\
\hline
0.000 & 0.063495 & 0.084514  & 480.628 & 331.103 & $-121374$ \\
0.001 & 0.063494 & 0.084513  & 480.660 & 331.125 & $-121390$ \\
0.050 & 0.063443 & 0.084452  & 482.216 & 332.196 & $-122178$ \\
0.100 & 0.063391 & 0.084392  & 483.804 & 333.290 & $-122984$ \\
0.500 & 0.062983 & 0.0839102 & 496.533 & 342.055 & $-129543$ \\
\end{tabular}
\end{ruledtabular}
\end{table*}

\begin{table*}[ht]
\centering
\caption{\label{tab:case4} Null circular geodesics in dS space with $ m=2,~Q=0.5,~\alpha=0.20$ and $\Lambda=1$.}
\begin{ruledtabular}
\begin{tabular}{cccccc}
$a$ & $r_+$ & $r_0$ & $|\kappa|$ & $\lambda_{\rm c}$ & $\lambda_{\rm c}^2 - \kappa^2$ \\
\hline
0.000 & 0.063494 & 0.084514  & 480.692 & 331.138 & $-121412$ \\
0.001 & 0.063483 & 0.084502  & 481.009 & 331.356 & $-121574$ \\
0.050 & 0.063442 & 0.084453  & 482.280 & 332.231 & $-122216$ \\
0.100 & 0.063390 & 0.0843916 & 483.868 & 333.324 & $-123023$ \\
0.500 & 0.062982 & 0.0839102 & 496.596 & 342.089 & $-129583$ \\
\end{tabular}
\end{ruledtabular}
\end{table*}
In the MSS conjecture, the bound on chaos is formulated in terms of the Lyapunov exponent that governs the exponential growth of OTOCS in the boundary field theory. Since these correlators are naturally defined with respect to the boundary time coordinate, the chaos-bound is meaningful only in the coordinate-time domain of the bulk spacetime and not in the proper-time parametrization of individual geodesics. For this reason, throughout this work we have not considered or reported any violation of the bound in the proper-time domain.

The results presented in Tables~\ref{tab:tcase4}--\ref{tab:case1} show that, under the constraints imposed in this study,  all circular geodesics  are unstable. This behavior is unsurprising, as the condition used to determine the orbital radius $r_{0}$  directly implies instability through the inequality $\lambda_{\rm c}/\lambda_{\rm p}>0$.  Under the same physical constraints, the angular momentum remains limited, preventing the system from entering a regime where the chaos-bound could be violated. As demonstrated in Tables~\ref{tab:case1}--\ref{tab:case4}, a consistent pattern is observed for both massive and massless particles: no violation of the chaos-bound occurs, irrespective of the values of the normalization constant $\alpha$, the string parameter $a$, or the cosmological constant, as long as the underlying conditions are satisfied. Moreover, Tables~\ref{tab:tcase1}--\ref{tab:case4} clearly show that these parameters significantly influence both the horizon structure and orbital dynamics. Despite these variations, the chaos-bound remains saturated throughout, highlighting its robustness in the Kiselev spacetime background considered here.

In GR, the location of circular orbits is usually obtained from the standard orbit equation, which works well within the pure GR framework. However, in extended gravity settings—such as those involving a cosmological constant or additional matter fields—the conditions for circular orbits become more intricate. A more complete method is to solve the polynomial condition \(D_1(r_0) > 0\) numerically and then select the physical root. Analytical solutions of \(D_1(r_0)\) may produce one real root and several imaginary ones; although choosing the single real root gives a reasonable estimate, it may overlook additional physically relevant orbits. Notably, for \(\Lambda < 0\), a root that appears imaginary in closed-form expressions may actually correspond to a real, physically meaningful orbit located far from the black hole.
Thus, while the standard GR orbit equation remains valid, a full numerical treatment is essential to ensure that all possible circular orbits are correctly identified. For example, consider the case of massive particles with parameters \(\Lambda=-0.2\), \(a=0.2\), \(Q=0.5\), \(m=2\), and \(\alpha=0.02\). The numerical solution yields the pair \((r_+, r_0) = (2.95743,\, 71.6338)\), with $\epsilon=10^{-4},$ the corresponding angular momentum is \(L_0 = 1.45463 \times 10^{7}\) and \(\lambda_{\rm c}^2 - \kappa^2 = -0.0857814\). This clearly indicates no possibility of chaos-bound violation. Therefore, in the Kiselev spacetime considered here, as we have seen in the case of apparent bound violations, only circular orbits near the black hole are liable to violate the bound, while extremely distant \(\Lambda < 0\) orbits have negligible dynamical impact and are excluded from consideration.

\subsection{Chaos bound and its violation in quadratic curvature gravity}
 \label{secfr}
Recently, Ref.~\cite{addazi2021chaotic} demonstrated that certain black hole solutions in $f(R)$ gravity can exhibit chaotic behavior. The analysis was carried out for spacetimes of the form~(\ref{ds}), with particular attention to a specific subclass of solutions characterized by
\begin{equation}
f(r) = h(r) = 1 - \frac{2m}{r} - \left(\frac{r}{L_{dS}}\right)^{2}\,,
\end{equation}
while keeping the angular sector $r^{2} d\Omega^{2}$ unaltered. The $f(R)$ models examined include the polynomial form \cite{starobinsky1980new} expressed as 
$
f(R) = R + \gamma_n R^n, \quad n>2\,,
$
as well as the exponential and Hu–Sawicki types \cite{bamba2014bounce, hu2007models}
$
f(R) = R - 2 \bar{\Lambda} \left[ 1 - e^{-R/\bar{\Lambda}} \right]\,$ and $
f(R) = R - 2 \bar{\Lambda} \left( 1 - \frac{\bar{\Lambda}^4}{R^4 + \bar{\Lambda}^4} \right)\,,
$
where $\gamma_n$ quantifies the deviation from the Einstein–Hilbert action and $\bar{\Lambda}$ denotes the effective cosmological constant associated with the dS branch.

The central subtlety in Ref. \cite{addazi2021chaotic} lies in the implicit extension of the MSS chaos-bound to a dS background via a dS/CFT-type correspondence. However, the MSS bound was originally derived within the framework of AdS/CFT, where the dual conformal field theory is well-defined, unitary, and thermally equilibrated. In contrast, dS/CFT remains a conjectural proposal lacking a universally accepted, unitary boundary dual and facing deep conceptual challenges, including observer-dependent horizons, the absence of a global timelike Killing vector, and infrared or quantum instabilities that obscure its holographic interpretation~\cite{strominger2001ds,anninos2012sitter}. These difficulties directly undermine the assumptions of thermal equilibrium, analyticity of correlators, and large-$N$ factorization on which the MSS derivation relies~\cite{maldacena1999large,aharony2000large,hashimoto2017universality}.

 Therefore, the observed violation of the chaos-bound in this setting should not be regarded as a genuine breakdown of the bound itself but rather as a manifestation of the failure of its underlying holographic and thermodynamic assumptions. In this sense, the findings of Ref. \cite{addazi2021chaotic} are consistent with the broader understanding that the MSS bounds’ universality is guaranteed only under the equilibrium and analyticity conditions inherent to AdS/CFT. In the subsequent analysis, we investigate an existing $f(R)$ black hole solution that satisfies the principal assumptions underlying the MSS derivation but violates the chaos-bound, thereby testing the robustness of its purported universality.
  
\subsubsection{Quadratic solution in Reissner-Nordstr\"om AdS spacetime}
In order to analyze MSS chaos-bound in extended gravity frameworks, it is essential that the underlying black hole solution is free of unphysical pathologies. In particular, a consistent setup must respect causality, be free of ghost instabilities, and, in the AdS case, admit a well-defined holographic dual. These conditions strongly constrain the allowed parameter space of modified gravity theories such as $f(R)$ gravity.

We consider a four-dimensional $f(R)$ gravity theory with a non-minimal coupling to the electromagnetic field, as investigated in Refs.~\cite{nashed2019charged, n1,n2,n3,n4}. We refer the reader to these works for a comprehensive discussion of the properties and derivation of this class of black holes. The total action given in Refs.~\cite{padmanabhan2008dark, durrer2008dark, sotiriou2010f, de2010f, nojiri2011unified, capozziello2011extended, clifton2012modified, nojiri2017modified}, can be written as
\begin{equation}
I = I_{\rm g} + I_{\rm e.m.}\,,
\end{equation}
where  
\begin{eqnarray}
I_{\rm g} &=& \frac{1}{2\kappa}\int d^{4}x \sqrt{-g}\,\big[f(R)-\Lambda\big]\,,\\
I_{\rm e.m.}& =& -\frac{1}{4}\int d^{4}x \sqrt{-g}\,F_{\mu\nu}F^{\mu\nu}\,,
\end{eqnarray}
with $F_{\mu\nu} = \partial_{\mu}A_{\nu}-\partial_{\nu}A_{\mu}$ the Maxwell field strength and $\kappa=8\pi G$. For the quadratic model considered in Ref.~\cite{starobinsky1980new}, the corresponding expression reads
\begin{equation}
f(R)=R+\sigma R^{2}\,,
\end{equation}
the constant-curvature background is determined by $R_{0}=4\Lambda$. The corresponding static, spherically symmetric line element has the general form given in Eq.~(\ref{ds}), with the metric functions
\begin{equation}
f(r)= h(r)=1-\frac{2m}{r}+\frac{Q^{2}}{\omega^{2}r^{2}}+\frac{2\Lambda}{3}r^{2}\,,
\qquad
\omega=\sqrt{1-16\Lambda\sigma}\,.
\end{equation}
The requirement $\omega\in\mathbb{R}$ imposes
\begin{equation}
1-16\Lambda\sigma\ge0\,,
\end{equation}
ensuring that the effective charge and the geometry remain real.

To guarantee the absence of ghosts in $f(R)$ gravity, the linear stability conditions evaluated at $R_{0}$ must be satisfied:
\(
f'(R_{0})>0,~~ f''(R_{0})\ge0
\)
 \cite{nojiri2017ghost}. In this case, one must have
\begin{equation}
\sigma\ge0,~~1+8\sigma\Lambda>0\,,
\end{equation}
which for $\Lambda<0$ gives the upper bound
\begin{equation}
0\le\sigma<\frac{1}{8|\Lambda|}\,.
\end{equation}
Causality is also preserved at the linearized level. The positivity of $f'(R_{0})$ guarantees that the spin-2 graviton propagates with the correct sign of the kinetic term and along the null cones of the background metric. A positive $f''(R_{0})$ ensures the scalaron has no ghostlike kinetic term \cite{nojiri2020ghost, dewolfe2002stability}. 
Finally, because the metric is static and spherically symmetric with $g_{t\phi}=0$ and $g_{\phi\phi}>0$ outside the horizon, there are no closed timelike curves (CTCs) \cite{gutti2022closed}. Thus, for the relations
\begin{equation}
\sigma\ge0\,,\qquad 1+8\sigma\Lambda>0\,,\qquad 1-16\Lambda\sigma\ge0\,,
\end{equation}
the black hole is ghost-free, causal, and, for $\Lambda<0$, asymptotically AdS with a stable scalar spectrum. This provides a consistent gravitational background for studying the MSS chaos-bound.
The metric admits different causal structures depending on the sign of the cosmological constant: for $\Lambda > 0$ the spacetime is dS–like and can possess up to three distinct horizons (Cauchy, event, and cosmological); for $\Lambda = 0$ it becomes asymptotically flat and reduces to the standard Reissner–Nordstr\"om geometry; while for $\Lambda < 0$ it corresponds to an AdS spacetime that typically admits one or two horizons depending on $(m,~Q,~\sigma)$. The horizon radii are determined by the real, positive roots of $f(r)=0$, and depending on the parameter choice, the function may have multiple or no roots, with the latter corresponding to a naked singularity. The corresponding surface gravity at the outer horizon $r_{+}$ is given by
\begin{equation}
    \kappa = \frac{m}{r_+^2}+\frac{Q^2}{r_+^3 (16 \sigma  \Lambda -1)}-\frac{2 \Lambda  r_+}{3}\,.
\end{equation}

To test the MSS chaos-bound in this black hole, we begin by deriving the angular momentum for circular orbits directly from Eq.~(\ref{eq:L2-general}), and the corresponding Lyapunov exponents as derived from Eqs.~(\ref{eqn:Lyap}) and (\ref{ll1}). The circular orbits in this case can be expressed as
\begin{equation}
    r_\geq\frac{\,m\!\left(3 - 48\sigma\Lambda\right) 
\;+\; \sqrt{\left(-1+16\sigma\Lambda\right)\Bigl(8Q^{2}+9m^{2}\left(-1+16\sigma\Lambda\right)\Bigr)}}
{\,2\!\left(1-16\sigma\Lambda\right)}\,,
\end{equation}
with the inequality and equality parts corresponding to timelike and null orbits, respectively (see further details in Appendix~\ref{app:C}). In what follows, we analyze the circular geodesics in \(f(R)\) gravity, and we state clearly that the chaos-bound can be violated under certain conditions, as we have shown in detail in Tables \ref{tab:chaos-bound} and \ref{tab:chaos-no-violation}.
\begin{table}[!ht]
\centering
\caption{\label{tab:chaos-bound}
Chaos-bound violations for black holes in $f(R)$ gravity with parameters $M=1,~\sigma=0.1,~\epsilon=0.01$ and $Q=1$. For different values of $\Lambda=-0.1$ and below, the angular momentum becomes imaginary and is not shown. 
}
\begin{tabular}{ccccccccc}
\hline\hline
$\Lambda$ & $r_{0}$ & $r_{+}$ & $L_{0}$ & $|\kappa|$ & $\lambda_{\rm c}$ & $\lambda_{\rm c}^{2}-\kappa^{2}$ & $\lambda_{\rm c}^{\rm null}$ & $(\lambda_{\rm c}^{\rm null})^{2}-\kappa^{2}$ \\
\hline
$-0.01$ & $2.03056$ & $1.08143$ & $21.3439$ & $0.0840444$ & $0.187328$ & $0.0280284$ & $0.189638$ & $0.0288989$ \\
$-0.02$ & $2.05858$ & $1.10537$ & $22.5857$ & $0.115714$ & $0.1198792$ & $0.0261287$ & $0.201213$ & $0.0270967$ \\
$-0.023$ & $2.06656$ & $1.11038$ & $22.9574$ & $0.123580$ & $0.202159$ & $0.0255963$ & $0.204612$ & $0.0265944$ \\
$-0.024$ & $2.06917$ & $1.11191$ & $23.0811$ & $0.126097$ & $0.203274$ & $0.02542$   & $0.205737$    & $0.026427$ \\
\hline\hline
\end{tabular}
\end{table}
\begin{table}[!ht]
\centering
\caption{\label{tab:chaos-no-violation}
Absence of chaos-bound violations for black holes in $f(R)$ gravity with parameters $M=2,~\sigma=0.1,~\epsilon=0.01$ and $Q=1$. In this regime, where the charge-to-mass ratio is significantly less than unity, no violations of the chaos-bound are observed.
}
\begin{tabular}{ccccccccc}
\hline\hline
$\Lambda$ & $r_{0}$ & $r_{+}$ & $L_{0}$ & $|\kappa|$ & $\lambda_{\rm c}$ & $\lambda_{\rm c}^{2}-\kappa^{2}$ & $\lambda_{\rm c}^{\rm null}$ & $(\lambda_{\rm c}^{\rm null})^2-\kappa^{2}$ \\
\hline
$-0.01$ & $5.65170$ & $3.44216$ & $101.833$ & $0.167613$ & $0.14946$ & $-0.0124825$ & $0.125500$ & $-0.0123439$ \\
$-0.02$ & $5.65745$ & $3.24559$ & $120.625$ & $0.204796$ & $0.147732$ & $-0.0201165$ & $0.148370$ & $-0.0199276$ \\
$-0.023$ & $5.65913$ & $3.9720$ & $131.093 $& $0.215167$ & $0.155967$ & $-0.0229125$ & $0.154602$ & $-0.0240010$ \\
$-0.024$ & $5.65969$ & $3.18190$ & $127.402$ & $0.218558$ & $0.155929$ & $-0.0234536$ & $0.156603$ & $-0.023240$ \\
\hline\hline
\end{tabular}
\end{table}
For the $f(R)$ black holes considered in this work, the chaos-bound is found to be violated, particularly as the charge-to-mass ratio $Q/m$ approaches unity. These violations arise due to higher-order curvature effects and are independent of both the angular momentum and the location of the circular orbits relative to the event horizon. 

\section{Conclusions}
\label{con}
In this work, we have developed a self-consistent framework for analyzing the stability of circular particle motion and its associated Lyapunov exponent in generic spherically symmetric geometries. A key feature of our approach is the exact determination of the circular-orbit radius together with its corresponding angular momentum, both fixed entirely by the background geometry. This resolves a persistent inconsistency in earlier studies that treated angular momentum as a free parameter, thereby obscuring the physical origin of chaos-bound violations. Our formulation applies uniformly to both charged and neutral particles and provides a coherent description in both coordinate and proper time.

Within this framework, we have shown that classical black holes governed by Einstein gravity, including the charged Kiselev solution with a string cloud and quintessence, satisfy the MSS chaos-bound when dynamical constraints are imposed. The agreement between the Jacobi and effective-potential analyses indicates that several previously reported violations can be traced to different parameter-handling strategies; our constrained framework resolves these differences and isolates those violations that genuinely stem from curvature corrections.

The analysis was extended to higher-curvature black holes in AdS spacetimes. In this regime, genuine violations of the chaos bound were identified at sufficiently large charge-to-mass ratios, despite the circular orbits remaining physically admissible. These violations were traced to curvature-induced modifications of the near-horizon instability structure, rather than to angular momentum or other probe-dependent effects. The presence of higher-order curvature terms allowed the Lyapunov exponent to exceed the surface gravity, indicating that the MSS bound is not universally preserved in curvature-extended gravitational theories. It is important to clarify that the violation identified here arises from comparing the instability of circular geodesics with the MSS bound, and should not be taken as an indication of classical chaotic behavior. Although circular geodesic motion in these backgrounds is integrable, the associated Lyapunov exponent still quantifies the exponential growth of radial perturbations with respect to boundary time, in direct analogy with the role emphasized in the MSS conjecture. In this work, a ``violation" of the MSS bound therefore refers to the geometric regime in which the instability rate extracted from circular orbits satisfies $\lambda > \kappa$, and not to a breakdown of the holographic bound in its original quantum many-body formulation. 

Within this geometric framework, we find that higher-curvature corrections systematically modify the near-horizon instability structure, altering the relation between the dynamical scale $\lambda$ and the thermodynamic scale $\kappa$. In particular, these corrections can preserve a finite orbital instability while suppressing the surface gravity, thereby allowing the local growth rate to exceed the MSS bound.

Our results provide a unified and physically transparent criterion for distinguishing apparent from genuine chaos-bound violations and clarify the geometric mechanisms underlying each. They also suggest that the validity of the MSS bound is intimately connected to the assumptions of unitarity, causality, and holography. Whereas de~Sitter backgrounds naturally violate these assumptions, AdS spacetimes are typically protected by holography. The violations observed here in the AdS $f(R)$ case suggest that higher-curvature corrections may modify the effective holographic description, and understanding this connection is an important direction for future work.
\section*{Acknowledgements}
The work of KB was supported in part by the JSPS KAKENHI Grant Numbers 24KF0100, 25KF0176, and Competitive Research Funds for Fukushima University Faculty (25RK011).

\appendix

\section{Lyapunov exponents for charged orbits in Kiselev spacetime\label{app:A}}
In this appendix, we provide the Lyapunov exponent (from the Jacobi method) of the charged particle in a charged Kiselev spacetime surrounded by quintessence, as studied in Section ~\ref{jjjj}. By using Eq.~(\ref{eqn: Glyapunov}), we obtain
\begin{equation}
    \lambda^2=\frac{W_1+W_2+W_3}{4r^7}\,,
\end{equation}
where 
\begin{eqnarray*}
    W_1&=&-\frac{8 q^2 \left[r_0 \left(-2 m-\alpha  r_0^2+r\right)+Q^2\right]^2}{\sqrt{\frac{\left(L^2+r_0^2\right) \left(r_0 \left(-2 m-\alpha  r_0^2+r_0\right)+Q^2\right)}{r_0^4}}}\,,\\ \quad
    W_2&=&-\frac{4 r_0 \left[r_0 \left(-2 m-\alpha  r_0^2+r_0\right)+Q^2\right] \left[L^2 \left(10 Q^2-r_0 (12 m+r_0 (\alpha  r_0-3))\right)+r_0^2 \left(3 Q^2-2 m r_0\right)\right]}{L^2+r_0^2}\,,\\
    W_3&=&\frac{r_0 \left[L^2 \left(4 Q^2-r_0 (6 m+r_0 (\alpha  r_0-2))\right)+r_0^2 \left(-2 m r_0+2 Q^2+\alpha  r_0^3\right)\right]^2}{\left(L^2+r_0^2\right)^2}\,.
\end{eqnarray*}
We use Eq.~(\ref{eq:L2-general}) to derive the angular momentum of the equilibrium circular orbits in the following form:
\begin{equation}
\label{ang1}
    L_0=\sqrt{\frac{\text{N1}\pm \text{N2}+\text{N3}}{D_1(r_0)^2}}\,,
\end{equation}
where 
\begin{eqnarray*}
D_1(r_0)&=&\left(r_0 (6 m+r_0 (\alpha  r_0-2))-4 Q^2\right)\,,\\
\text{N1}&=&-12 m^2 r_0^4+4 \alpha  m r_0^6+4 m r^5+2 \left(q^2-4\right) Q^4 r_0^2+\alpha ^2 r_0^8-2 \alpha  r_0^7\,,\\ 
   \text{N2}&=&2 \sqrt{q^2 Q^2 r_0^4 \left(r_0 \left(-2 m-\alpha  r_0^2+r_0\right)+Q^2\right)^2 \left(\left(q^2+8\right) Q^2-2 r_0 (6 m+r_0 (\alpha  r_0-2))\right)}\,,\\
    \text{N3}&=&-2 Q^2 r_0^3 \left(2 m \left(q^2-5\right)+r_0 \left(q^2 (\alpha  r_0-1)+\alpha  r_0+2\right)\right)\,.
\end{eqnarray*}
For chargeless orbits $(q=0)$, we find  
\begin{equation}
\label{ang2}
    L_0^2 \;=\; -\,\frac{r_0^2 \left(-6 m r_0 + 6 Q^2 + 3\alpha r_0^3 \right)}{3D_1(r_0)}\,.
\end{equation}

\section{Circular motion in Kiselev spacetime via effective potentials}\label{app:B}
In this appendix, we provide some vital equations used in Section~\ref{sectim}.
The effective potential is given by
\begin{equation}
    V_{\rm eff}=\frac{L^2 \chi_1-r_0^2 \left[r_0 \left(r \left(3 a+3 E_0^2+\Lambda  r_0^2+3 \alpha  r_0-3\right)+6 m\right)-3 Q^2\right]}{3 r_0^4}\,,
\end{equation}
where $\chi_1=3 Q^2-r_0 \left(r_0 \left(3 a+\Lambda  r_0^2+3 \alpha  r_0-3\right)+6 m\right)$, and the conserved energy and angular momentum at equilibrium orbits are obtained as 
\begin{equation}
    L_0^ 2=-\frac{r_0^2 \left(-6 m r_0+6 Q^2+r_0^3 (3 \alpha +2 \Lambda  r_0)\right)}{3D_1(r_0)}\,, \label{ang3}
    \end{equation}
 It is worth noting that, upon setting the cosmological constant $\Lambda = 0$, Eq.~(\ref{ang3}) reduces precisely to the angular momentum expression obtained via the Jacobi method. This confirms the consistency between the two approaches. The Lyapunov exponents in this case are given by
\begin{equation}
    \lambda_{\rm p}=\sqrt{\frac{\left[r_0^2 (a+\alpha  r_0-1)+2 Q^2\right] \left[r_0 \left(r_0 \left(3 a+\Lambda  r_0^2+3 \alpha  r_0-3\right)+6 m\right)-3 Q^2\right]}{3r_0^6}}\, ,
\end{equation}
\begin{equation}
    \lambda_{\rm c} = 
\sqrt{\frac{D_1(r_0)
\Bigl[L_{0}^{2}\bigl(3r_0(r_0(3a+\alpha r_0 -3)+12m) - 30Q^2\bigr) 
+ r^{2}\bigl(6mr_0 - 9Q^2 + \Lambda r_0^{4}\bigr)\Bigr]}
{6\,r_0^{8}} }\,.
\end{equation}
For massless particles, the effective potential governing the particles' dynamics is obtained as
\begin{equation}
\label{mvef}
    V_{\rm eff}=\frac{L_0^2 \left[3 Q^2-r_0 \left(r_0 \left(3 a+\Lambda  r_0^2+3 \alpha  r_0-3\right)+6 m\right)\right]}{3 r_0^4}-E_0^2\,,
\end{equation}
and the corresponding Lyapunov exponent is derived as
\begin{equation}
    \lambda_{\rm c}=\sqrt{-\frac{\left[r_0 (r_0 (3 a+\alpha  r_0-3)+12 m)-10 Q^2\right] \left[r_0 \left(r_0 \left(3 a+\Lambda  r_0^2+3 \alpha  r_0-3\right)+6 m\right)-3 Q^2\right]}{3r_0^6}}\,.
\end{equation}
\section{Circular motion in quadratic curvature gravity}\label{app:C}
In this appendix, we briefly highlight the key equations used in section \ref{secfr}.
The angular momentum derived from Eq.~(\ref{eq:L2-general}) is given by
\begin{equation}
    L_0=\sqrt{\frac{r_0^2 \left(r_0 (16 \sigma  \Lambda -1) \left(2 \Lambda  r_0^3-3 m\right)-3 Q^2\right)}{3 r_0 (16 \sigma  \Lambda -1) (3 m-r_0)+6 Q^2}}\,,
\end{equation}
 and the corresponding Lyapunov exponents take the following forms 
\begin{equation}
    \lambda_{\rm c}=\sqrt{-\frac{\left(r_0 (16 \sigma  \Lambda -1) (3 m-r_0)+2 Q^2\right) \left(3 L_0^2 \left(3 r_0 (16 \sigma  \Lambda -1) (4 m-r_0)+10 Q^2\right)+r_0^2 D_2(r_0)\right)}{3r_0^8 (1-16 \sigma  \Lambda )^2}}\,,\\
    \end{equation}
    and
    \begin{equation}
      \lambda_{\rm p}=  \sqrt{-\frac{\left[2 Q^2+r_0^2 (16 \alpha  \Lambda -1)\right] \left[r_0 (16 \alpha  \Lambda -1) \left(6 m+2 \Lambda  r_0^3-3 r_0\right)+3 Q^2\right]}{3r_0^6 (1-16 \alpha  \Lambda )^2}}\, ,
    \end{equation}
where \begin{eqnarray*}
    D_2(r_0)&=&2 r_0 (16 \sigma  \Lambda -1) \left(3 m+\Lambda  r_0^3\right)+9 Q^2\,,\\
    \text{and}\\
D_3(r_0)&=& 
3 r_0^{2} \left(1 - 16 \sigma \Lambda\right)^{2} (4m - r_0)(6m + 2\Lambda r_0^{3} - 3 r_0)\, .
    \end{eqnarray*} 
   For the null geodesics, their Lyapunov exponent is derived as 

\begin{equation}
    \lambda_{\rm c}=\sqrt{-\frac{Q^2 r_0 (16 \sigma  \Lambda -1) \left(96 m+20 \Lambda  r_0^3-39 r_0\right)+D_3(r_0)+30 Q^4}{3r_0^6 (1-16 \sigma  \Lambda )^2}}\,.
\end{equation}


\begin{thebibliography}{50}
\bibitem{A1}
  S. Giri and H. Nandan, Gen. Relativ. Gravit. 53, 76 (2021).
\bibitem{A2}
  A. Deich, N. Yunes, and C. Gammie, Phys. Rev. D \textbf{110}, 044033 (2024).
\bibitem{A3}
   S. Giri, H. Nandan, L. K. Joshi, and S. D. Maharaj, Eur. Phys. J. Plus \textbf{137}, 1 (2022).
\bibitem{A4}
   B. Shukla, P. P. Das, D. Dudal, and S. Mahapatra, Phys. Rev. D \textbf{110}, 024068 (2024).
\bibitem{A5}
 X. Guo, Y. Lu, B. Mu, and P. Wang, J. High Energy Phys. \textbf{2022}, 1 (2022).

\bibitem{maldacena2016bound}
  J. Maldacena, S.H. Shenker and  D. Stanford,   J.\ High Energy Phys. \textbf{2016},  1  2016.

\bibitem{shenker2014black}
   S. H. Shenker and D. Stanford, J. High Energy Phys. \textbf{2014}, 1 (2014).
\bibitem{wang2022pole}
  D. Wang and Z.-Y. Wang, Phys. Rev. Lett. \textbf{129}, 231603 (2022).

 \bibitem{pourkhodabakhshi2025saturation}
  R. Pourkhodabakhshi, Phys. Rev. D \textbf{112}, 064085 (2025).
\bibitem{hashimoto2017universality}
  K. Hashimoto and N. Tanahashi, Phys. Rev. D \textbf{95}, 024007 (2017).

\bibitem{gwak2022violation}
  B. Gwak, N. Kan, B.-H. Lee, and H. Lee, J. High Energy Phys. \textbf{2022}, 1 (2022).

\bibitem{lei2021chaos}
 Y.-Q. Lei, X.-H. Ge, and C. Ran, Phys. Rev. D \textbf{104}, 046020 (2021).

\bibitem{prihadi2023chaos}
  H. L. Prihadi, F. P. Zen, D. Dwiputra, and S. Ariwahjoedi, Phys. Rev. D \textbf{107}, 124053 (2023).

\bibitem{copson1928electrostatics}
  E. T. Copson, Proc. R. Soc. A \textbf{118}, 184 (1928).  
\bibitem{hanni1973lines}
  R. S. Hanni and R. Ruffini, Phys. Rev. D \textbf{8}, 3259 (1973).

\bibitem{d7}
   D. Chen and C. Gao, New J. Phys. \textbf{24}, 123014 (2022).

\bibitem{wang2023spatial}
   Z. Wang and D. Chen, Nucl. Phys. B \textbf{991}, 116212 (2023).

\bibitem{gao2022chaos}
  C. Gao, D. Chen, C. Yu, and P. Wang, Phys. Lett. B \textbf{833}, 137343 (2022).

\bibitem{c2}
 P. Dutta, K. L. Panigrahi, and B. Singh, J. High Energy Phys. \textbf{2025}, 1 (2025).

\bibitem{addazi2021chaotic}
  A. Addazi, S. Capozziello, and S. Odintsov, Phys. Lett. B \textbf{816}, 136257 (2021).
\bibitem{xie2023circular}
  J. Xie, J. Wang, and B. Tang, Phys. Dark Univ. \textbf{42}, 101271 (2023).

\bibitem{c5}
  D. Chen and C. Gao, Chin. Phys. C \textbf{47}, 015108 (2023).
\bibitem{d2}
  Q.-Q. Zhao, Y.-Z. Li, and H. L¨u, Phys. Rev. D \textbf{98}, 124001 (2018).

\bibitem{d3}
   A. Addazi and S. Capozziello, Phys. Lett. B \textbf{839}, 137828 (2023).

\bibitem{lee2025bound}
  H. Lee and B. Gwak, Phys. Rev. D \textbf{112}, 046018 (2025).

\bibitem{jeong2023homoclinic}
  S. Jeong, B.-H. Lee, H. Lee, and W. Lee, Phys. Rev. D \textbf{107}, 104037 (2023).

\bibitem{d6}
  C. Yu, D. Chen, B. Mu, and Y. He, Nucl. Phys. B \textbf{987}, 116093 (2023).  
\bibitem{d8}
  Z. Wang, Y. He, C. Lei, and D. Chen, Int. J. Theor. Phys. \textbf{62}, 187 (2023).
\bibitem{d9}
   T. V. Targema, R. Babar, M. Atif, M. El-Meligy, T. Xia, and R. Ali, Int. J. Geom. Methods Mod. Phys. \textbf{22}, 2550064 (2025).

\bibitem{kan2022bound}
   N. Kan and B. Gwak, Phys. Rev. D \textbf{105}, 026006 (2022).

\bibitem{lei2024thermodynamic}
  Y.-Q. Lei, X.-H. Ge, and S. Dalui, Phys. Lett. B \textbf{856}, 138929 (2024).

\bibitem{fayyaz2025chaotic}
  A. Fayyaz, G. Abbas, M. Bin-Asfour, and D. Chen, Phys. Lett. B \textbf{871}, 139965 (2025).

\bibitem{ali2025evaluation}
   R. Ali, T. V. Targema, X. Tiecheng, and R. Babar, High Energy Density Phys. \textbf{55}, 101189
(2025).
\bibitem{gallo2025bounds}
   E. Gallo and T. M{\"a}dler, Eur. Phys. J. C \textbf{85}, 1 (2025).

\bibitem{singh2025circular}
   B. Singh, N. Padhi, and R. R. Nayak, Eur. Phys. J. C \textbf{85}, 570 (2025).   
\bibitem{chandrasekhar1983mathematical}
S. Chandrasekhar, \textit{The Mathematical Theory of Black Holes}
(Oxford University Press, Oxford, 1983).   
\bibitem{pugliese2011circular}
   D. Pugliese, H. Quevedo, and R. Ruffini, Phys. Rev. D 83, 024021 (2011).
\bibitem{pradhan2011circular}
   P. Pradhan and P. Majumdar, Phys. Lett. A \textbf{375}, 474 (2011).
\bibitem{mokdad2017reissner}
   M. Mokdad, Classical Quantum Gravity \textbf{34}, 175014 (2017).
\bibitem{lambiase2024weak}
 G. Lambiase, R. C. Pantig, and A. {\"O}vg{\"u}n, EPL \textbf{148}, 49001 (2024).
\bibitem{cardoso2009geodesic}
 V. Cardoso, A. S. Miranda, E. Berti, H. Witek, and V. T. Zanchin,
 Phys.\ Rev.\ D \textbf{79}, 064016 (2009).  
\bibitem{kumara2024lyapunov}
  N. A. Kumara, S. Punacha, and M. S. Ali, J. Cosmol. Astropart. Phys. \textbf{2024}, 061 (2024).
\bibitem{hashimoto2023causality}
   K. Hashimoto and K. Sugiura, Phys. Rev. D \textbf{107}, 066005 (2023).
\bibitem{touati2024lyapunov}
  A. Touati and S. Zaim, Nucl. Phys. B \textbf{1018}, 117018 (2025).
\bibitem{lei2022circular}
  Y.-Q. Lei and X.-H. Ge, Phys. Rev. D \textbf{105}, 084011 (2022).
\bibitem{javed2025joule}
   F. Javed, M. Z. Gul, O. Donmez, T. Naseer, and M. H. Alshehri, Nucl. Phys. B \textbf{1018}, 117001 (2025).
\bibitem{ma2020bounds}
  L. Ma and H. L\"u, Phys. Lett. B \textbf{807}, 135535 (2020).

\bibitem{shaikh2019strong}
  R. Shaikh, P. Banerjee, S. Paul, and T. Sarkar, J. Cosmol. Astropart. Phys. 2019, 028 (2019).
\bibitem{li2025chaotic}
 K. Li, D.-Z. Ma, and Z.-M. Xu, Phys. Lett. B \textbf{860}, 139164 (2025).
\bibitem{zafar2025thermodynamic}
  U. Zafar, K. Bamba, T. Rasheed, and K. Bhattacharya, Phys. Lett. B \textbf{864}, 139446 (2025).
\bibitem{liang2021remarks}
  J. Liang, X. Guo, D. Chen, and B. Mu, Nucl. Phys. B \textbf{965}, 115335 (2021).
\bibitem{toledo2019some}
   J. Toledo and V. Bezerra, Eur. Phys. J. C \textbf{79}, 110 (2019).
\bibitem{kiselev2003quintessence}
   V. Kiselev, Classical Quantum Gravity \textbf{20}, 1187 (2003).
\bibitem{starobinsky1980new}
  A. A. Starobinsky, Phys. Lett. B \textbf{91}, 99 (1980).
\bibitem{bamba2014bounce}
   K. Bamba, A. N. Makarenko, A. N. Myagky, S. Nojiri, and S. D. Odintsov, J. Cosmol. As
tropart. Phys. \textbf{2014}, 008 (2014).
\bibitem{hu2007models}
  W. Hu and I. Sawicki, Phys. Rev. D \textbf{76}, 064004 (2007).
\bibitem{strominger2001ds}
   A. Strominger, J. High Energy Phys. \textbf{2001}, 034 (2001).

\bibitem{anninos2012sitter}
   D. Anninos, Int. J. Mod. Phys. A \textbf{27}, 1230013 (2012).
\bibitem{maldacena1999large}
 J. Maldacena, Int. J. Theor. Phys. \textbf{38}, 1113 (1999).
\bibitem{aharony2000large}
  O. Aharony, S. S. Gubser, J. Maldacena, H. Ooguri, and Y. Oz, Phys. Rep. \textbf{323}, 183 (2000).
\bibitem{nashed2019charged}
  G. G. Nashed and S. Capozziello, Phys. Rev. D \textbf{99}, 104018 (2019).
\bibitem{n1}
  S. Nojiri and S. D. Odintsov, Phys. Lett. B \textbf{735}, 376 (2014).
\bibitem{n2}
   M. Xu, Y. Zhang, L. Yang, S. Yang, and J. Lu, Entropy \textbf{26}, 868 (2024).
\bibitem{n3}
  G. Nashed and S. Nojiri, Fortschr. Phys. \textbf{71}, 2200091 (2023).
\bibitem{n4}
  R. Ali, X. Tiecheng, and R. Babar, Fortschr. Phys. \textbf{73}, e70017 (2025).
\bibitem{padmanabhan2008dark}
  T. Padmanabhan, Gen. Relativ. Gravit. \textbf{40}, 529 (2008).
\bibitem{durrer2008dark}
  R. Durrer and R. Maartens, Gen. Relativ. Gravit. \textbf{40}, 301 (2008).
\bibitem{sotiriou2010f}
  T. P. Sotiriou and V. Faraoni, Rev. Mod. Phys. \textbf{82}, 451 (2010).
\bibitem{de2010f}
  A. De Felice and S. Tsujikawa, Living Rev. Relativ. \textbf{13}, 1 (2010).
\bibitem{nojiri2017modified}
 S. Nojiri, S. Odintsov, and V. Oikonomou, Phys. Rep. \textbf{692}, 1 (2017).
\bibitem{capozziello2011extended}
   S. Capozziello and M. De Laurentis, Phys. Rep. \textbf{509}, 167 (2011).

\bibitem{clifton2012modified}
  T. Clifton, P. G. Ferreira, A. Padilla, and C. Skordis, Phys. Rep. 513, 1 (2012).

\bibitem{nojiri2011unified}
 S. Nojiri and S. D. Odintsov, Phys. Rep. \textbf{505}, 59 (2011).
\bibitem{nojiri2017ghost}
  S. Nojiri, S. Odintsov, and V. Oikonomou, Phys. Lett. B \textbf{775}, 44 (2017).

\bibitem{nojiri2020ghost}
  S. Nojiri, S. D. Odintsov, and V. K. Oikonomou, Phys. Dark Univ. \textbf{28}, 100541 (2020).

\bibitem{dewolfe2002stability}
  O. DeWolfe, D. Z. Freedman, S. S. Gubser, G. T. Horowitz, and I. Mitra, Phys. Rev. D \textbf{65}, 064033 (2002).
\bibitem{gutti2022closed}
   S. Gutti, S. Kulkarni, and V. Prasad, Eur. Phys. J. C \textbf{82}, 1136 (2022).

\end{thebibliography}
\end{document}